\documentclass{amsart} 
\usepackage{amssymb}
\usepackage{amsmath}
\usepackage[mathscr]{eucal}
\usepackage{color}
\usepackage[dvips]{graphicx}
\usepackage{enumerate}

\newtheorem{thm}{Theorem}[section]
\newtheorem{prop}{Proposition}[section]  
\newtheorem{lem}{Lemma}[section]
\newtheorem{rem}{Remark}[section]
\newtheorem{cor}{Corollary}[section]

\numberwithin{equation}{section}

 \newcommand{\Zd}{{\mathbb Z}^d}
  \newcommand{\Zh}{{\mathbb Z}_h}
   \newcommand{\Zhd}{{\mathbb Z}_h^d}
  \newcommand{\Rd}{{\mathbb R}^d}
 \newcommand{\Thd}{{\mathbb T}_{\!1\!/h}^d}
 
  \newcommand{\Rtw}{{\mathbb R}^2}
    \newcommand{\Ctw}{{\mathbb C}^2}
     \newcommand{\Ztw}{{\mathbb Z}^2}
     \newcommand{\Zhtw}{{\mathbb Z}_h^2}
  \newcommand{\Thtw}{{\mathbb T}_{\!1/h}^2} 
 \newcommand{\Ttw}{{\mathbb T}^2}

 \newcommand{\sumZd}{\sum_{n\in \Zd}}
  \newcommand{\sumZtw}{\sum_{n\in \Ztw}}
 
  \newcommand{\charnh}{{\chi}_{{}_{I_{n,h}} } }
   \newcommand{\Inh}{ I_{\! n\!,\!\,h}}
  
   \newcommand{\Fh}{{\mathcal F}_h}
    \newcommand{\iFh}{{\overline{\mathcal F}}_h}
     \newcommand{\Fc}{\mathcal F}
     \newcommand{\iFc}{\overline{\mathcal F}}
     
 \newcommand{\Dh}{{\mathbb D}_{m,h}}   
   \newcommand{\wDh}{\widehat{\mathbb D}_{m,h}}
    \newcommand{\Dc}{{\mathbb D}_{m} }
 \newcommand{\wDc}{\widehat{\mathbb D}_{m}}

\begin{document}

\title[DISCRETE DIRAC OPERATORS ON 2D LATTICES]{Continuum limits for discrete Dirac operators on 2D square lattices}

\author{Karl Michael Schmidt}
\address{Karl Michael Schmidt: School of Mathematics,  Cardiff University, Senghennydd Road,
Cardiff CF24 4AG, Wales, UK}
\email{schmidtkm@cardiff.ac.uk}
\author{Tomio Umeda}
\address{Tomio Umeda: Department of
Mathematical Sciences,  University of Hyogo, Himeji 671-2201,
Japan}
\email{	umeda@sci.u-hyogo.ac.jp}

\thanks{
Corresponding author: Karl Michael Schmidt, email: SchmidtKM@cardiff.ac.uk \\
T.Umeda was partially supported by
 the Japan Society for the Promotion of Science
     ``Grant-in-Aid for Scientific Research'' (C)
    No.  26400175.  This author thanks the Erwin Schr\"odinger International Institute for Mathematics and
    Physics, University of Vienna, for the support during the workshop on 
    \lq\lq Spectral Theory of Differential Operators in Quantum Theory\rq\rq,
    November 7 -- 11, 2022, organized by J. Behrndt, F. Gesztesy, A. Laptev and C. Tretter.
\\
The authors have no competing interests to declare that are relevant to the content of this article. }

\date{\today}

\maketitle


\smallskip


\begin{abstract}
We discuss the continuum limit of discrete Dirac operators on the square  lattice  in $\mathbb R^2$ as 
the mesh size tends to zero.  To  this end, we propose the most natural
and  simplest embedding  of $\ell^2(\mathbb Z_h^d)$ into $L^2(\mathbb R^d)$,
which  enables us 
to compare the discrete Dirac operators with the continuum Dirac operators
in the same Hilbert space $L^2(\mathbb R^2)^2$.
 In particular, we prove that the discrete Dirac operators 
  converge to the continuum Dirac operators
  in the strong resolvent sense.
Potentials are assumed to be
bounded and uniformly continuous functions on $\mathbb R^2$
and allowed to be  complex matrix-valued. 
We also prove that the discrete Dirac operators do not converge to the continuum Dirac operators in the norm
resolvent sense. This is closely related to the observation that the Liouville theorem does not hold in discrete
complex analysis.
\end{abstract}


\section{Introduction}

This paper is concerned with 
the discrete Dirac operator  ${\mathbb D}_{m,h}+ V_h$   defined by
\begin{equation}\label{eq:d-dirac2-1}
{\mathbb D}_{m,h} + V_h= 
\begin{pmatrix}
m &  i \partial_{1,h}^* + \partial_{2,h}^* \\
- i \partial_{1,h} + \partial_{2,h}  & -m
\end{pmatrix}
+ V_h
\text{ \;  in  \;}
\ell^2(\Zh^2)^2,
\end{equation}
which is a discrete
analogue of the two-dimensional Dirac operator defined by      
\begin{equation} \label{eq:dirac2}
{\mathbb D}_{m} + V=  -i \sigma_1 \frac{\partial}{\partial x_1} 
-i\sigma_2  \frac{\partial}{\partial x_2} 
+ m \sigma_3 + V(x)
\;\text{  in }
L^2(\mathbb R^2)^2,
\end{equation}
where $m\ge 0$ and $\sigma_1$, $\sigma_2$, $\sigma_3$ are  the Pauli matrices, 
\begin{equation}\label{eqn:1-3}
\sigma_1 =
\begin{pmatrix}
0&1 \\ 1& 0
\end{pmatrix}, \,\,\,
\sigma_2 =
\begin{pmatrix}
0& -i  \\ i&0
\end{pmatrix}, \,\,\,
\sigma_3 =
\begin{pmatrix}
1&0 \\ 0&-1
\end{pmatrix},
\end{equation}
and  $V$  is a complex matrix valued function.
For the  definition  of  $\ell^2(\Zh^2)^2$, 
see  (\ref{eq:ell2}) in  section \ref{sec:2};
 the finite difference  operators $\partial_{j,h}$ and $\partial_{j,h}^*$  ($j \in \{ 1, \, 2 \}$)
are defined in  (\ref{eq:diff-op11})
and (\ref{eq:diff-op21}) in section \ref{sec:2},
respectively;
 for  $V_h$, 
see    (\ref{eq:V0}) in section \ref{sec:main1}.

 We remark that  both operators    ${\mathbb D}_{m,h}$ in 
 (\ref{eq:d-dirac2-1}) and ${\mathbb D}_{m}$ in (\ref{eq:dirac2}) possess  supersymmetry  structure
 (see \cite[Chapter 5]{Thaller}, \cite[Chapter 3]{Tretter}).
  The discrete Dirac operator (\ref{eq:d-dirac2-1}) can be rewritten 
in  a form analogous to  (\ref{eq:dirac2}),
\begin{equation*}
{\mathbb D}_{m,h} = 
-i\sigma_1
\begin{pmatrix}
 \partial_{1,h}&  0 \\
0 & - \partial_{1,h}^* 
\end{pmatrix}
-i\sigma_2
\begin{pmatrix}
 \partial_{2,h}&  0 \\
0 & - \partial_{2,h}^* 
\end{pmatrix}
+
m\sigma_3 + V_h.
\end{equation*}

It is widely recognized that 2D Dirac operators, especially in the massless case,
 have been  the object of   extensive research  in the context of  graphene since 
 its discovery in 2004,  see \cite{CGPNG} or \cite{P} for an exposition. 
 In particular, we would like to mention the work 
  \cite{NG}, which reported that electron transport in graphene is  essentially governed  by a massless Dirac  equation 
  and that a variety 
  of unusual phenomena  are characteristic of two-dimensional  Dirac fermions. 
  These are the main reasons why we focus on the two-dimensional case, 
  although 
  it is apparent that the methods and ideas to be developed below in the present paper 
  are  directly applicable to the one-dimensional and the three-dimensional cases. 
The discussions in these two cases will appear elsewhere.

It is natural to make an attempt to show that the discrete operator (\ref{eq:d-dirac2-1})
converges to the continuum operator (\ref{eq:dirac2})
as the mesh size $h$ of the lattice  $\Zh^2$ tends to $0$. 
However, there is a difficulty 
 in that  these two operators work in completely different Hilbert spaces. 
 For example, it is not immediately obvious  how one can make sense of the expression 
$({\mathbb D}_{m,h} + V_h) - ({\mathbb D}_{m} + V)$. 
For this reason, 
it is necessary to embed $\ell^2(\Zhtw)^2$
onto an appropriate subspace of $L^2(\Rtw)^2$.

In this paper, we propose  a simple and  natural 
embedding of
$\ell^2(\Zhd)$
into $L^2(\Rd)$ by assigning to each element in 
$\ell^2(\Zhd)$
a step function in $L^2(\Rd)$:
\begin{equation} \label{eq:myemdding}
[J_h f](x) := \sumZd f(hn)  \,{\chi}_{{}_{I_{n,h}} } \!(x)   \qquad (f \in \ell^2(\Zh^d))
\end{equation}
where $\charnh$ is a characteristic function of the set       
\begin{equation*}
I_{n,h}:= \{ x \, | \, h n_j \le x_j  < h(n_j +1),  \  j\in\{1,\cdots,d\}  \}
\end{equation*}
 (see  subsection \ref{sec:embed}). 
 We find it is important  
 that
the discrete Fourier transform can be naturally defined for $J_h f$  (see  subsection \ref{sec:dFI}). 
Also,  the use of  
step functions is  desirable
 from the  point of view  of numerical analysis.
  This idea of embedding 
$\ell^2(\Zhd)$
into $L^2(\Rd)$
 induces a subspace
$L^2(\Zhd)$ of $L^2(\Rd)$.
With this embedding, one can naturally define the difference operators  
$\partial_{j,h}$ and $\partial_{j,h}^*$ in 
$L^2(\Zhd)$,  
the subspace of step functions of the form (\ref{eq:myemdding})
(cf. subsection \ref{sec:embed}), 
and hence the discrete Dirac operators ${\mathbb D}_{m,h}$ in $L^2(\Zhtw)^2$.
For the reasons mentioned here, 
the discrete Dirac operators
${\mathbb D}_{m,h}$ in $L^2(\Zhtw)^2$ 
 are the exact counterparts of 
 the discrete Dirac operators ${\mathbb D}_{m,h}$
in $\ell^2(\Zhtw)^2$,  so we can identify these two operators.
In other words,
we are able to 
regard the discrete Dirac operators ${\mathbb D}_{m,h} + V_h$ as
an operator acting in $L^2(\mathbb R^2)^2$
with domain $L^2(\Zh^d)^2$, 
and able to compare the discrete Dirac operators ${\mathbb D}_{m,h} + V_h$  with
 the  continuum  Dirac operators ${\mathbb D}_{m} + V$
 in the same Hilbert space $L^2(\mathbb R^2)^2$.
 The purpose of the present paper is to show, 
 with the embedding  operator 
 defined by (\ref{eq:myemdding}), 
 that the resolvents of the discrete Dirac operators 
(\ref{eq:d-dirac2-1})
converge to the continuum 
Dirac operators (\ref{eq:dirac2}) in the strong resolvent sense as the mesh
size $h$ tends to $0$ 
(see Theorem \ref{thm:x} in section \ref{sec:main0} and 
Theorem \ref{thm:V} in section \ref{sec:main1}).
In addition, we show that the discrete operator $\mathbb D_{m,h}$ does not
converge to the continuum operator $\mathbb D_m$ in the norm resolvent sense (see Theorem \ref{thm:nonoco} 
 in section \ref{sec:main0}).
As a motivation for the proof of the latter theorem, we observe that the Liouville theorem does not hold in discrete
complex analysis (see Remark \ref{rem:monodiff} in section \ref{sec:main0}).


In connection with the embedding operator (\ref{eq:myemdding}), 
we would like to mention the works by \cite{CGJ} and \cite{NT}, in which 
the embedding operators are defined by
\begin{equation}\label{eq:distortedemdding1}
\sum_{n\in \Zd} \rho \big( (x - hn)/h \big) f(hn)   \qquad (f \in \ell^2(\Zh^d)),
\end{equation}
with  $\rho$   a smooth and (possibly rapidly) decreasing function.  
However, in this paper, we do not adopt this type of
embedding of $\ell^2(\Zhd)$
into $L^2(\Rd)$, because of the following  three 
reasons.
Firstly, the embedding operator (\ref{eq:distortedemdding1}) depends on the choice of the function $\rho$.
Secondly, the   embedded 
functions defined by (\ref{eq:distortedemdding1}) are  smooth, and no longer
 discrete objects. In fact,  the  discrete Fourier transform is not applicable to smooth functions.
Thirdly,
 the difference operators working on  smooth functions
 can be regarded as a mixture of discreteness and continuum, and 
 may not 
 be regarded as the exact counterparts of difference operators in  $\ell^2(\Zhtw)$. 
On the other hand, 
as was  pointed out in \cite{CGJ},  
it is inevitable to introduce the embedding operator 
(\ref{eq:distortedemdding1}) 
and 
a modification of the  discrete Dirac operators in $\ell^2(\Zhtw)^2$
if one would like to show 
the norm resolvent convergence. 

We should like to remark that if one replaces the function $\rho$ with the characteristic
function ${\chi}_{{}_{I_{0,1}} }$ 
(i.e., ${\chi}_{{}_{I_{n,h}} }$ with $n=0$, $h=1$)
then the embedding operator (\ref{eq:distortedemdding1}) coincides with
our embedding operator
 (\ref{eq:myemdding}).

 When we apply  our idea to  the  discrete Dirac operator (\ref{eq:d-dirac2-1})
 to discuss  the continuum limit 
 as the mesh size $h$ tends to $0$,
 we  require  a convergence theorem  for  the orthogonal projection $P_h$ 
 onto 
 the closed subspace
 $L^2(\Zh^2)^2$  of $L^2(\Rtw)^2$.
Also, we  
require a  discrete Fourier transform on
$L^2(\Zh^2)$  (which is 
essentially the same, 
but not identical to the Fourier series with coefficients in $\ell^2(\Zh^2)$), 
and need to prove  a convergence theorem  for  the
discrete Fourier transform  as $h \to 0$. 
 Indeed, we will establish  both convergence theorems 
 in the strong topology of
$L^2(\Rd)$.
  Precise descriptions  are  given in subsection 
\ref{sec:dFI} and section \ref{sec:discreteFT}.
With these strong convergence theorems, we can prove that the resolvents of the discrete Dirac operator 
 (\ref{eq:d-dirac2-1}) strongly converge to those
of the Dirac operator  (\ref{eq:dirac2}) 
 in  $L^2(\Rtw)^2$.

In the literature, there have been few papers  studying  spectral properties of discrete Dirac operators on 2D or 3D lattices,
while there have been many working on 1D lattices,
 see for example the recent works  \cite{A}, \cite{BMT}, \cite{CIKS}, \cite{CY},
\cite{KMRS}, \cite{K}, \cite{KT}, \cite{PC}, \cite{RS}, \cite{S1}, \cite{S2}.
 We mention in passing that the discrete Dirac operator on a 1D lattice, when written in
matrix form, is a tri-diagonal matrix and indeed unitarily equivalent to a discrete Schr\"odinger-type operator;
for example, with the unitary operator $U : \ell^2(\mathbb Z_h) \rightarrow \ell^2(\mathbb Z_h)^2$ given as
$(U f)(nh) = (-1)^n \begin{pmatrix} f(2nh) \\ f((2n+1)h) \end{pmatrix}$, 
\begin{equation*}
 U^* \begin{pmatrix} m & \partial_h^* \\ \partial_h & -m \end{pmatrix} U = h \left(-\partial_h^* \partial_h + q \right),
\end{equation*}
where $q(nh) = (-1)^n m/h - 2/h^2$.

 To our knowledge, 
 the only papers working on discrete Dirac operators in dimensions 2 and 3 are \cite{CGJ} and  \cite{POC}.
 The lack of works on the continuum limit of 
 discrete analogs of 
quantum Hamiltonians, as far as we know, is
hardly surprising in view of the
fact that research on this topic began rather recently;  see \cite{CGJ}, \cite{IJ} and \cite{NT}.

 Finally,
 we would like to mention yet another idea of natural embedding.
Indeed, we find  the embedding in \cite{IJ} is natural in the sense that it assigs to each element in 
$\ell^2(\Zhd)$
a discrete object in  ${\mathcal S}^{\prime}(\Rd)$ and discrete Fourier transform is naturally
associated.
With this embedding operator,
continuum limits of lattice Schr\"odinger
operators for various models  were  investigated in \cite{IJ}. 
In particular, lattice Laplacians satisfying suitable assumptions 
were shown to converge to the 2D Dirac operators (\ref{eq:dirac2}) with $m=0$ and $V=0$.
 Specifically for the hexagonal (graphene) lattice, see also \cite{FW}.

 The present paper is organised as follows.  Section 2 illustrates the idea of embedding 
$\ell^2(\Zhd)$ into $L^2(\Rd)$ and  shows how the  finite difference operators
in the embedded space can  naturally  be defined.  
It also describes how the discrete Fourier transform can be
extended as an operator
in $L^2(\Rd)$.
In section 3, convergence of discrete Fourier transform in $L^2(\Rd)$ is discussed.
Resolvent  convergence of the discrete Dirac operator
without potential  is discussed in section 4,
based on the results
obtained 
in the previous sections.
Strong resolvent convergence of the discrete Dirac operators
with potentials
is discussed in section 5.

 \section{Embedding of $\ell^2(\Zhd)$ into $L^2(\Rd)$ and discrete Fourier transform in $L^2(\Rd)$}
 \label{sec:2}

In applications to the Dirac operator (\ref{eq:d-dirac2-1}),  the underlying Hilbert space is $\ell^2(\Zhtw)^2 := \ell^2(\Zhtw)\otimes\Ctw$, 
which consists of $\Ctw$-valued functions on the $2$-dimensional lattice.
In this section, we focus on the space of complex-valued functions on the $d$-dimensional lattice,
$\ell^2(\Zhd)$, and related spaces of functions on $\Rd$; the results naturally extend to the corresponding spaces of $\Ctw$-valued functions.

The $d$-dimensional square  lattice with the mesh size $h>0$ is denoted by
\begin{equation*}
\Zhd:=\{ \, hn \, \big| \, n \in \Zd  \}.
\end{equation*}
 The Hilbert space 
\begin{align}\label{eq:ell2}
{}\ell^2({\mathbb Z}_h^d):= \big\{ \, f \, \big|  \,  f : {\mathbb Z}_h^d \to \mathbb C, \;\;
\sum_{n\in \Zd} | f(hn)|^2 < \infty \, \big\},   
\end{align}
has the standard inner product 
\begin{equation}
\big( f, \, g \big)_{\ell^2({\mathbb Z}_h^d)}=
\sum_{n\in \Zd} f(hn)   \, \overline{g(hn)} .
\end{equation}
For  $f\in \ell^2(\Zhd)$, define  
 \begin{equation}\label{eq:diff-op11}
  [\partial_{j,h} f](hn)
  := \frac{1}{h}\Big\{ f(h(n+e_j) )-  f(hn)  \Big\}  \quad ( j \in \{1, \dots, d\} ),
\end{equation}
where $ 
  e_1= (1, \,  0,\, \dots, 0), \dots ,  e_d= ( 0,\, \dots, 0, \, 1)$. 
 The adjoint  of  $\partial_{j,h}$ is 
  given by
 \begin{equation}\label{eq:diff-op21}
  [\partial_{j,h}^* f](hn)
  := \frac{1}{h}\Big\{ f(h(n-e_j)) -  f(hn)  \Big\}  \quad ( j \in \{1, \dots, d\} ),
\end{equation}
 so 
\begin{equation}\label{eq:ajoint1}
\big( \partial_{j,h}f, \, g \big)_{\ell^2({\mathbb Z}_h^d)}
=\big( f, \,  \partial_{j,h}^* g \big)_{\ell^2({\mathbb Z}_h^d)}, 
\quad  \forall f, g \in \ell^2(\Zhd).
\end{equation}
%

\vspace{5pt}  
\subsection{Embedding of $\ell^2(\Zhd)$ into $L^2(\Rd)$}\label{sec:embed}

We introduce  an embedding of $\ell^2(\Zhd)$ into $L^2(\Rd)$ by assigning
to $f \in \ell^2({\mathbb Z}_h^d)$ the step function 
\begin{equation} \label{eq:embedJ}
[J_h f](x) := \sumZd f(hn)  \,{\chi}_{{}_{I_{n,h}} } \!(x) \in L^2(\Rd),
\end{equation}
where $\charnh$ is a characteristic function of the set       
\begin{equation}
I_{n,h}:= \{ x \, | \, h n_j \le x_j  < h(n_j +1),   \  j\in\{1,\cdots,d\}  \};
\end{equation}
 clearly,
$\displaystyle\sumZd \charnh  \equiv 1$. 
Note that  
\begin{equation}\label{eq:embed0}
\Vert J_h f \Vert_{L^2(\Rd)} =  h^{d/2}  \Vert f \Vert_{\ell^2(\Zhd)},
\end{equation}
so  $h^{-d/2}J_h$ is an isometry from $\ell^2(\Zhd)$ into $L^2(\Rd)$.
Since 
$\ell^2(\Zhd)$ is a Hilbert space, 
 the image $J_h \big(  \ell^2(\Zhd) \big)$ is
a  closed subspace of the Hilbert space
$L^2(\Rd)$. 
We thus have an orthogonal decomposition
\begin{equation} \label{eq:embed1}
L^2(\Rd)= L^2(\Zhd) \oplus L^2(\Zhd)^{\perp}, \qquad  L^2(\Zhd):=J_h \big(  \ell^2(\Zhd) \big).
\end{equation}
(The notation $L^2(\Zhd)$ already appeared in \cite{IJ}, but  there it essentially denotes
$\ell^2(\Zhd)$; for details, see \cite[Subsection 2.1]{IJ}.)
The orthogonal projection $P_h$ of $L^2(\Rd)$ onto $L^2(\Zhd)$
can be described, for general $\varphi\in L^2(\Rd)$, as
$P_h \varphi = J_h \tilde\varphi$, where
\begin{equation}\label{eq:embed2}
\tilde\varphi(hn) = \frac 1 {h^d} \int_{I_{n,h}} \varphi (x) \, dx \qquad (n \in \mathbb Z^d).
\end{equation}

\vspace{10pt}
\begin{rem} \label{rem:expl0}
The following example illustrates the action of the projection $P_h$. Let $\varphi\in L^2(\mathbb R)$ be 
defined by $\varphi(x)=0 \  (|x|\ge 2h)$, \, $\varphi(x) = x + 2h \  (-2h < x < -h)$,
\, $\varphi(x) = h \ (|x|\le h)$, 
and $\varphi(x) = -x +2h \  (h < x <2h)$.
This function can be decomposed as $\varphi=f_h + g_h$ with
$$
f_h(x)=\left\{\begin{array}{ll}
  h & \mbox{if}\ -h \le x < h, \\
  \frac h 2 & \mbox{if}\ -2h \le x < -h\ \mbox{or}\ h \le x < 2h, \\
  0 & \mbox{otherwise},
  \end{array}\right. 
$$
and
$$
  g_h(x)=\left\{\begin{array}{ll}
  x+\frac{3h}2 & \mbox{if}\  -2h \le x < -h, \\
  -x + \frac{3h}2 & \mbox{if}\  h \le x < 2h , \\
   0 & \ \mbox{otherwise.}
  \end{array}\right.
$$
It is easy to see that $f_h\in L^2(\mathbb Z_h)$ and
$g_h\in L^2(\mathbb Z_h)^\bot$, so
$P_h \varphi=f_h$. 
\end{rem}

 As the above embedding gives a one-to-one relationship between the elements of $\ell^2(\Zhd)$ and of $L^2(\Zhd)$,
we can define the finite difference operators $\partial_{j,h}$ and $\partial_{j,h}^* $ on $L^2(\Zhd)$ by
applying them, as defined in (\ref{eq:diff-op11}) and (\ref{eq:diff-op21}), to the corresponding 
element of $\ell^2(\Zhd)$, such that
\begin{equation}\label{eq:diffops}
\partial_{j,h} J_h[f] := J_h[\partial_{j,h}f] \qquad
\partial_{j,h}^* J_h[f] := J_h[\partial_{j,h}^*f] \qquad (f\in\ell^2(\Zhd)).
\end{equation}
Then we again have
\begin{equation}\label{eq:ajoint2}
\big( \partial_{j,h}f, \, g \big)_{L^2({\mathbb Z}_h^d)}
=\big( f, \,  \partial_{j,h}^* g \big)_{L^2({\mathbb Z}_h^d)}
\qquad  (f, g \in L^2(\Zhd)).
\end{equation}
in analogy to (\ref{eq:ajoint1}).

\vspace{5pt}  
\subsection{Discrete Fourier transform} \label{sec:dFI}
 Let
 \begin{equation}\label{eq:htorus}
 \Thd= [ -\pi/h, \, \pi/h]^d = \{\xi\in\Rd \mid |\xi|_\infty \le \pi/h\},
 \end{equation}
where
$|\xi|_\infty = \max \{|\xi_1|, \dots, |\xi_d|\}$.
(Although we use a notation alluding to the interpretation, natural in the following, of this set as a flat
$d$-dimensional torus of side length $2\pi/h$, we emphasize that it is a bounded interval in $\Rd$.)
As the functions
\begin{equation*}
e_n(\xi) = \left(\frac h {2\pi}\right)^{d/2}\,e^{-ih n\cdot\xi} \qquad
(\xi\in\Thd; n\in\mathbb Z^d)
\end{equation*}
form an orthonormal basis of $L^2(\Thd)$, any function $f\in\ell^2(\Zhd)$ serves as a collection of Fourier
coefficients for a $d$-dimensional Fourier series in $L^2(\Thd)$,
\begin{equation*}
 \sum_{n\in\Zd} f(hn)\,e_n(\xi) \qquad (\xi\in\Thd).
\end{equation*}
In view of the bijection between $\ell^2(\Zhd)$ and $L^2(\Zhd)$, this motivates the following definition of
a {\em discrete Fourier transform}
$\Fh : L^2(\Zhd) \rightarrow L^2(\Thd)$,
\begin{align}
[\Fh J_h f](\xi) &:= h^{d/2} \sum_{n\in\Zd} f(hn) e_n(\xi) \nonumber \\
&=   (2\pi)^{-d/2} \int_{{\mathbb R}^d} \sumZd e^{-ihn\cdot \xi} f(hn)  \,\charnh (x) \, dx  
  \quad  (\xi \in \Thd) 
\label{eq:dfourier0}
 \end{align}
for $f\in\ell^2(\Zhd)$.
By Parseval's identity for the orthonormal basis $\{e_n \mid n\in\Zd\}$,
\begin{equation}
 \|\Fh J_h f\|_{L^2(\Thd)}^2 = h^d \sum_{n\in\Zd} |f(hn)|^2 = h^d\,\|f\|_{\ell^2(\Zhd)}^2 = \|J_h f\|_{L^2(\Zhd)}^2
\end{equation}
for any $f\in\ell^2(\Zhd)$, so $\Fh$ is a unitary operator.

Its inverse is the operator
$\iFh : L^2(\Thd) \rightarrow L^2(\Zhd)$,
 \begin{equation}\label{eq:idfourier0}
 [\iFh u](x) =
  \sumZd   \Big\{  \! (2\pi)^{-d/2}  \! \int_{\Thd} \!  e^{ihn\cdot \xi} \,  u(\xi) \, d\xi  
  \Big\} \charnh  \!(x) \quad (x\in\Rd).
\end{equation}
for $u\in L^2(\Thd)$.

 By direct computations, we have   
 \begin{equation}\label{eq:dfourier1}
[ \Fh  (\partial_{j,h}f) ](\xi) = \frac{1}{h}(e^{ih\xi_j} -1)[\Fh f](\xi)  \qquad (f \in L^2(\Zhd))
 \end{equation}
 and
 \begin{equation}\label{eq:dfourier1-1}
[ \Fh  (\partial_{j,h}^*f) ](\xi) = \frac{1}{h}(e^{-ih\xi_j} -1)[\Fh f](\xi)  \qquad (f \in L^2(\Zhd)).
 \end{equation}

Extending functions by 0 outside $\Thd$, the space $L^2(\Thd)$ naturally forms a closed subspace of $L^2(\Rd)$;
it is the range of
 the orthogonal projection $Q_{1\!/h}$ defined as the operator of multiplication with the characteristic function of $\Thd$.

Using the projections $P_h \in \mathbf B( L^2(\Rd))$ and 
$Q_{\!1\!/h} \in \mathbf B( L^2(\Rd))$, we  can 
extend $\Fh$  and its inverse $\iFh$
 to become elements of  $\mathbf B(L^2(\Rd))$ 
by setting
\begin{equation}\label{eq:dfourierL2}
\Fh:= Q_{\!1\!/h} \,\Fh P_h,  \quad \iFh:= P_h \iFh Q_{\!1\!/h}.
\end{equation}
Here $\mathbf B( L^2(\Rd))$ denotes the Banach space of all bounded linear operators in $L^2(\Rd)$, 
equipped with the uniform operator topology.
 Note  that $\Fh$ is a partial isometry from   $L^2({\mathbb R}^d)$ to $L^2({\mathbb R}^d)$ 
with the initial set $L^2(\Zhd)$
and 
the final set $L^2(\Thd)$, and that $\iFh$ 
is a partial isometry from  $L^2({\mathbb R}^d)$ to $L^2({\mathbb R}^d)$ 
with the initial set $L^2(\Thd)$
and 
the final set $L^2(\Zhd)$.
 Clearly,
\begin{equation}\label{eq:proj-dfour0}
\iFh \Fh = P_h,
\qquad \Fh \iFh = Q_{\!1\!/h},
\end{equation}
and 
\begin{equation}\label{eq:proj-dfour1}
P_h^{\perp} \iFh = \mathbf 0, \quad    Q_{\!1\!/h}^{\perp} \Fh = \mathbf 0.
\end{equation}
%

\vspace{10pt}
\section{Convergence of discrete Fourier transforms} \label{sec:discreteFT}

As in the previous section, we will work in $\mathbb C$-valued functions on $d$ dimensional Euclidean space $\Rd$.

For $\varphi \in \mathcal S(\Rd)$, the Schwartz space of  rapidly decreasing functions, 
we define
\begin{equation} \label{eq:phi_h}
\varphi_h(x)= \sumZd \varphi(hn) \charnh  \! (x).
\end{equation}
It is clear that $\varphi_h \in L^2(\Zhd)$ and $P_h \varphi_h = \varphi_h$.  However, we emphasize that  
$P_h \varphi\not= \varphi_h$ in general.
In fact, for each $h>0$ one can easily  choose a function
 $\varphi \in \mathcal S(\Rd)$ such that
 $(\varphi_h, \, \varphi-\varphi_h)_{L^2(\Rd)}\not=0$.

 \begin{rem} \label{rem:embed0}
 As was noted in (\ref{eq:embed0}),  we have
 $\Vert  \varphi_h \Vert_{L^2(\Rd)} = h^{d/2} \Vert \varphi_h \Vert_{\ell^2(\Zhd)}$.
 \end{rem}

\begin{lem}   \label{lem:convPh0}
Let $\varphi\in \mathcal S(\Rd)$ and $k \in \mathbb N$. 
\begin{list}{}{\itemindent=-15pt}
\item[\rm(i)] There exists a constant $C_{\!\varphi k}$, depending only on 
$\varphi$ and $k$, such that
\begin{equation} \label{eq:fhnfx-1}
| \varphi_h(x) - \varphi(x) | \le C_{\!\varphi k}\langle x \rangle^{-k} h   \qquad ( x \in \Rd, \, 0 < h < \frac{1}{\sqrt{d}}),
\end{equation}
where $\langle x \rangle=\sqrt{1+|x|^2}$.
In particular,  $\varphi_h$ is rapidly decreasing, i.e., $| \varphi_h(x)| \le C_{\varphi k}\langle x \rangle^{-k}$ for any $k \in \mathbb N$.
\item[\rm(ii)] There exists a constant $C_{\!\varphi}$, depending only on 
$\varphi$, such that 
\begin{equation}\label{eq:L2conv}
\Vert \varphi_h -\varphi \Vert_{L^2}\le C_{\!\varphi} h   \qquad ( 0 < h < \frac{1}{\sqrt{d}}).
\end{equation}
\end{list}
\end{lem}

\noindent
{\bf Proof.}
 We prove statement (i); then statement (ii) follows as a straightforward consequence, taking $k > d/2$.
Let $n\in\Zd$. Then, for $x\in I_{n,h}$,
\begin{equation*}
\varphi(hn) -\varphi(x)= (hn - x) \cdot  \! \int_0^1 \big(\nabla \varphi \big) (t(hn-x)+x) \, dt.
\end{equation*}
Consequently,
\begin{equation*}
\begin{aligned}
|\varphi(hn) -\varphi(x)| \, \langle x \rangle^{k}  &{} \\
&\hspace{-65pt}
\le  
| hn -x |   \int_0^1  
\frac{\langle x \rangle^{k} }{\langle t(hn-x)+x \rangle^k}
\big| \big(\nabla \varphi \big) (t(hn-x)+x) \big| \langle t(hn-x)+x \rangle^k dt  \\
&\hspace{-65pt}
\le 
\left(\sup_{y\in\Rd} \left| (\nabla\varphi)(y) \right| \langle y \rangle^k \right) \,| hn -x |   \int_0^1  
\frac{\langle x \rangle^{k} }{\langle t(hn-x)+x \rangle^k} dt.  
\end{aligned}
\end{equation*}
Now if $|x| \ge 2 \sqrt d h$, then
\begin{equation*}
|t(hn-x)+x| \ge |x| - |hn-x| \ge |x| - \sqrt d h \ge \frac{|x|}2,
\end{equation*}
so
\begin{equation*}
\langle t(hn-x)+x \rangle \ge \sqrt{1 + \frac{|x|^2}4} = \sqrt{1 + \frac{\langle x \rangle^2 - 1}4}
 \ge \frac{\langle x \rangle}2;
\end{equation*}
if $|x| < 2 \sqrt d h$, then (trivially) $\langle t(hn-x)+x \rangle \ge 1$ and
$\langle x \rangle < \sqrt{1 + 4 d h^2} < \sqrt 5$.
In either case,
\begin{equation*}
\int_0^1 \frac{\langle x \rangle^k}{\langle t(hn-x)+x \rangle^k}\,dt \le \sqrt 5^k,
\end{equation*}
and, noting that $|hn-x| \le \sqrt d h$, the inequality (\ref{eq:fhnfx-1}) follows. 
$\blacksquare$

\medskip
 The statement of Lemma \ref{lem:convPh0} (i) has the following immediate consequence. 

\begin{cor}
Let $\varphi\in \mathcal S(\Rd)$ and $k \in \mathbb N$. 
There exists a constant $C_{\!\varphi k}$, depending only on 
$\varphi$ and $k$, such that 
\begin{equation}\label{eq:fhnfx5}
\sumZd |\varphi(hn)|  \,\charnh(x) \le 
 |\varphi(x)| +  C_{\!\varphi k}  \, h\langle x \rangle^{-k}
  \qquad (x\in\Rd) 
\end{equation}
for  $0 < h < 1/\sqrt{d}$.
\end{cor}

\vspace{8pt}
In what follows, we shall use the notation
\begin{align}
\mathcal S_h^{\,\rm step}(\Rd)
&= 
\{ \;  \varphi_h=\sumZd \varphi(hn) \charnh   \; |  \; \varphi \in  \mathcal S(\Rd) \, \}, \\
\mathcal S_{0+}^{\,\rm step}(\Rd)
&=
 \bigcup_{h>0} \mathcal S_h^{\,\rm step}(\Rd).
\end{align}
 Note that, unlike $\mathcal S_h^{\,\rm step}(\Rd)$, the set of functions $\mathcal S_{0+}^{\,\rm step}(\Rd)$ is not a vector space. 
As we shall see in Lemma \ref{lem:prelim1} in subsection \ref{sec:convF1}, 
each $\varphi_h \in \mathcal S_{0+}^{\,\rm step}(\Rd)$ 
allows an explicit expression of its Fourier transform 
in a certain sense.

The following lemma is a direct consequence of Lemma \ref{lem:convPh0}(ii).

\begin{lem}   
$\mathcal S_{0+}^{\,\rm step}(\Rd)$ is a  dense  subset of $L^2(\Rd)$.
\end{lem}

 We can now prove the strong convergence of the orthogonal projectors $P_h$ and $Q_{1/h}$ to the identity.

\begin{lem}  \label{lem:Ph} 
For any $u\in L^2(\Rd)$,  $\Vert P_h u - u \Vert_{L^2(\Rd)} \to 0$ and $\Vert Q_{1/h} u - u \Vert_{L^2(\Rd)} \to 0$ as $h \to 0$. 
\end{lem}

\noindent
{\bf Proof.}
Since  $\Vert P_h \Vert_{\mathbf B(L^2(\Rd))}=1$ for all $h>0$ and
$\mathcal S(\Rd)$ is dense in $L^2(\mathbb R^d)$, 
it is sufficient to prove
that 
 for $\varphi \in \mathcal S(\mathbb R^d)$, $\Vert P_h \varphi - \varphi \Vert_{L^2(\Rd)} \to 0$ as $h \to 0$.

 Let $\varphi\in \mathcal S(\Rd)$  and, for each $h > 0$, let $\varphi_h \in L^2(\Zhd)$ be the step function defined in (\ref{eq:phi_h}).
Then $P_h \varphi_h = \varphi_h$, so
\begin{align*}
 \|P_h \varphi - \varphi\|_{L^2(\Rd)} &\le \|P_h \varphi - P_h \varphi_h\|_{L^2(\Rd)}+ \|\varphi_h - \varphi\|_{L^2(\Rd)} \\
 &\le \left(\|P_h\|_{\mathbf B(L^2(\Rd))} + 1\right)\, \|\varphi_h - \varphi\|_{L^2(\Rd)}\rightarrow 0 \qquad (h \rightarrow 0)
\end{align*}
by Lemma \ref{lem:convPh0}(ii).

Furthermore, for any $u\in L^2(\Rd)$
\begin{equation*}
\| Q_{1/h} u - u\|_{L^2(\Rd)}^2 = \int_{\Rd\setminus\Thd} |u|^2 \rightarrow 0 \qquad (h\rightarrow 0).
\end{equation*}
$\blacksquare$

 \vspace{10pt}

 \begin{rem}
 It is clear from the proof of Lemma \ref{lem:Ph} that 
 the projection $Q_{1/h}$ does not 
 converge to the identity operator  in the operator norm as $h \rightarrow 0$.
 The same is true for the projection $P_h$.
Indeed, for any $h > 0$ there is a function $\varphi \in L^2(\mathbb{R}^d)$ such that
$\|\varphi\|_{L^2(\mathbb{R}^d)} = 1$ and $\int_{I_{n,h}} \varphi = 0$ for all $n \in \mathbb{Z}^d$,
e.g.
\begin{equation*}
\varphi(x) = \sum_{n\in\mathbb{Z}^d} c_n \chi_{I_{n,h}}(x) \prod_{j=1}^d (x_j - h(n_j + \frac 1 2)) \qquad (x \in \mathbb{R}^d)
\end{equation*}
with suitable
$(c_n)_{n\in\mathbb{Z}^d}$.
Then, using  (\ref{eq:embed2}),
we find
\begin{equation*}
\|P_h \varphi - \varphi\|_{L^2(\mathbb{R}^d)} = \|0 - \varphi\|_{L^2(\mathbb{R}^d)} = 1,
\end{equation*}
so $\|P_h - I\|_{\mathbf{B}(L^2(\mathbb{R}^d))} \ge 1$ for any $h > 0$.
 \end{rem}

\vspace{5pt}  
\subsection{Convergence of $\Fh$ and $\iFh$}\label{sec:convF1}
 As usual, the Fourier transform $\Fc$ on $L^2(\Rd)$ and its inverse $\iFc$ arise by extension
of the integral operators
\begin{equation} \label{eq:fourier1}
[\Fc \varphi](\xi) = (2\pi)^{-d/2} \int_{\Rd} e^{-ix\cdot \xi} \varphi(x) \, dx \qquad (\xi\in\Rd; \varphi\in\mathcal S(\Rd))
\end{equation}
and
\begin{equation} \label{eq:ifourier1}
[\iFc \varphi](x) = (2\pi)^{-d/2} \int_{\Rd} e^{ix\cdot \xi} \varphi(\xi) \, d\xi \qquad (x\in\Rd; \varphi\in\mathcal S(\Rd)),
\end{equation}
respectively.
We emphasize that we can compare $\Fh$ with $\Fc$ and $\iFh$ with $\iFc$ on
$L^2(\Rd)$, as we have extended the discrete Fourier transform and its inverse to all of $L^2(\Rd)$
in (\ref{eq:dfourierL2}).

\begin{lem}  \label{lem:prelim1}  
Let $\varphi_h \in \mathcal S_{0+}^{\,\rm step}(\Rd)$.
Then
$$[\Fc \varphi_h] (\xi) =   
\Big\{
 (2\pi)^{-d/2} \sumZd  \varphi(hn)   \, h^d  
 e^{-ihn\cdot \xi}
 \Big\}
  \, \displaystyle{\prod_{j=1}^d} a(h\xi_j)  \qquad (\xi\in\Rd), $$
 where 
$$a(\theta)=\dfrac{1-e^{-i\theta}}{i\theta} \;\;\; (\theta \in \mathbb R).$$
\end{lem}

\noindent
{\bf Proof.}   
Let $n \in \Zd$ and $h>0$. 
 A direct computation shows that
\begin{equation}\label{eq:prod-a}
\int_{\Inh}  e^{-ix\cdot \xi}   dx 
= 
\prod_{j=1}^d  
e^{-ihn_j \xi_j} \frac{e^{-ih\xi_j}-1}{-i\xi_j}
=
h^d e^{-ihn\cdot \xi} 
\prod_{j=1}^d  a(h\xi_j).
\end{equation}
We then have
\begin{equation} \label{eq:fourier2}
\begin{split}
[ \Fc \varphi_h] (\xi)  
&=  
(2\pi)^{-d/2}   \sumZd  \varphi(hn) \int_{\Inh} e^{-ix\cdot \xi}  dx  \\
&=
  (2\pi)^{-d/2} \sumZd  \varphi(hn)   \, h^d  
 e^{-ihn\cdot \xi} 
\prod_{j=1}^d  a(h\xi_j). \\
  \end{split}
\end{equation}
This completes the proof. {$\blacksquare$}

\vspace{10pt}

An immediate consequence of (\ref{eq:dfourier0}) and Lemma \ref{lem:prelim1} is the following
corollary, which we expect will be useful from the view point of numerical analysis
of discrete approximations of Fourier transform.

\vspace{10pt}

\begin{cor}\label{cor:prelim1}
Let $\varphi_h \in \mathcal S_{0+}^{\,\rm step}(\Rd)$.
Then 
\begin{equation}\label{eq:fhnfx6}
[\Fh \varphi_h - \Fc \varphi_h] (\xi) =   \Big(  1 - \prod_{j=1}^d a(h\xi_j) \Big) [\Fh \varphi_h](\xi)
\qquad (\xi \in \Thd).
\end{equation}
\end{cor}

\vspace{1pt}
 \begin{rem} \label{rem:a(theta)}
Note that $a(\theta)\to 1$ as $\theta \to 0$, and that   $|a(\theta)|\le 1$ for all $\theta$. 
\end{rem}

\begin{rem} 
One can deduce that $\Fh \varphi$ converges locally in  $L^2$  to $\Fc \varphi$ for any $\varphi \in L^2(\Rd)$
in the following manner.

Since  $\Vert \Fh \Vert_{\mathbf B(L^2(\Rd))}=1$  for all $h>0$,  and
since  $\mathcal S(\Rd)$ is  dense  in $L^2(\mathbb R^d)$, 
it is sufficient to prove the local convergence in $L^2$ for $\varphi \in \mathcal S (\Rd)$.

Let $\varphi \in \mathcal S(\Rd)$ and 
let $\varphi_h$ be given by (\ref{eq:phi_h}).
 Then, by (\ref{eq:dfourier0}) and (\ref{eq:fhnfx5}),
\begin{equation} \label{eq:fhnfx7}
\begin{split}
|[\Fh \varphi_h](\xi)|
&=
  (2\pi)^{-d/2}\Big|
   \int_{{\mathbb R}^d} \sumZd e^{-ihn\cdot \xi} \varphi(hn)  \,\charnh \! (x) \, dx  \Big|   \\
&\le
 (2\pi)^{-d/2}
   \int_{{\mathbb R}^d} \sumZd  | \varphi(hn) | \,\charnh \! (x) \, dx   \\
&\le
(2\pi)^{-d/2}
   \int_{{\mathbb R}^d}  \big(
     |\varphi(x)| + {\rm const}_{\!\varphi} \, \langle x \rangle^{-d-1}
     \big) \, dx  =: C_{\varphi}  < \infty.
\end{split}
\end{equation}        
Taken together with Corollary \ref{cor:prelim1}, this estimate implies  that
$$
\Vert [\Fh - \Fc] \varphi_h \Vert_{L^2(K)} \le  C_{\varphi} 
 \sup_{\xi \in K} \Big| 1 - \prod_{j=1}^d a(h\xi_j) \Big| \to   0 \quad \text{ as } h \to 0
$$ 
for  any compact  subset $K$ of \,$\Rd$.    
The decomposition
\begin{equation}\label{eq:decomp-0}
[\Fh - \Fc] \varphi
=\Fh ( \varphi - \varphi_h) + [\Fh - \Fc] \varphi_h 
+ \Fc ( \varphi_h - \varphi) ,
\end{equation}
together with Lemma \ref{lem:convPh0} (ii),
gives the local convergence in $L^2$.  However, Lemma \ref{lem:Fourier} below
shows  a stronger convergence of $\Fh$.
\end{rem}

\vspace{10pt}
We close section \ref{sec:discreteFT} with the  (fairly straightforward) proofs of convergences of 
$\iFh$ and $\Fh$ respectively.

\begin{lem}   \label{lem:iFourier}
For any   $u \in L^2(\mathbb R^d)$,  $\Vert \iFh u  - \iFc u \Vert_{L^2(\Rd)}  \to 0$ as $h \to 0$.
\end{lem}

\noindent
{\bf Proof.} 
Since  $\Vert \iFh \Vert_{\mathbf B(L^2(\Rd))}=1$ for all $h>0$,  and
since  $C_0^{\infty}(\mathbb R^d)$ is dense in $L^2(\mathbb R^d)$, 
it is sufficient to prove
the assertion for $u \in C_0^{\infty}(\mathbb R^d)$.

Let $u \in C_0^{\infty}(\mathbb R^d)$  and choose $h_* >0$ so that
$\Thd \supset \text{supp} [u]$ for all $h \in (0, \, h_*)$. 
Then we have, by (\ref{eq:idfourier0}) and by the fact that $\Thd \supset \text{supp} [u]$,
\begin{align}
\iFh u  &= 
 \sumZd   \Big\{  \! (2\pi)^{-d/2}  \! \int_{\Rd} \!  e^{ihn\cdot \xi} \,  u(\xi) \, d\xi  
  \Big\} \charnh  \!   \nonumber \\ 
 {}&=
 \sumZd  \varphi(hn)  \charnh  \!    
   \end{align}
for all  $h \in (0, \, h_*)$, where we set $\varphi := \iFc u$.  This equality implies that
\begin{equation}\label{eq:fhnfx9}
 | \iFh u (x) - \iFc u (x) |^2 =   \sumZd  | \varphi(hn) - \varphi(x)  |^2\charnh  \!(x) 
  \qquad (x\in\Rd). 
\end{equation}
In view of the fact that $\varphi\in \mathcal S(\Rd)$, it follows from  (\ref{eq:fhnfx9}) and 
Lemma \ref{lem:convPh0}(i) that 
$\Vert \iFh u  - \iFc u  \Vert_{L^2(\Rd)}  \to 0$ as $h \to 0$.
$\blacksquare$

\vspace{8pt}
 \begin{lem}   \label{lem:Fourier}
 For any $\varphi \in L^2(\Rd)$,
\begin{equation}\label{eq:fhnfx10}
\Vert  \mathcal F_h \varphi - \mathcal F \varphi \Vert_{L^2(\Rd)} \to 0
\;\mbox{ as }\;  h \to 0.
\end{equation}
\end{lem}

\noindent
{\bf Proof.} 
We first prove that for any $u  \in L^2(\Rd)$
\begin{equation}\label{eq:fhnfx11}
\Vert  [\mathcal F_h  - \mathcal F \, ] \, \iFc  Q_{1/h} u  \Vert_{L^2(\Rd)} \to 0
\;\mbox{ as }\;  h \to 0.
\end{equation}
In fact, we see that the left hand side of (\ref{eq:fhnfx11}) is  equal to
\begin{equation}
\Vert  \mathcal F_h   \iFc  Q_{1/h} u -   Q_{1/h} u  \Vert_{L^2(\Rd)},
\end{equation}
which is bounded by
\begin{equation} \label{eq:fhnfx12}
\Vert     \iFc  Q_{1/h} u -   \iFh  u  \Vert_{L^2(\Rd)}.
\end{equation}
Here we have used the fact that  $Q_{\!1\!/h}=\Fh \iFh$ (recall (\ref{eq:proj-dfour0}),
and the fact that 
$\Vert \Fh \Vert_{\mathbf B(L^2(\Rd), L^2(\Rd)) }=1$.
Since  
\begin{equation*}
Q_{1/h} u= (Q_{1/h} u - u) +u,
\end{equation*}
(\ref{eq:fhnfx12}) is estimated by
$\Vert Q_{1/h} u - u \Vert_{L^2(\Rd)}+\Vert \iFc u  - \iFh u \Vert_{L^2(\Rd)}$.
The fact (\ref{eq:fhnfx11}) now follows from Lemmas \ref{lem:iFourier} and \ref{lem:Ph}. 

We next prove (\ref{eq:fhnfx10}). 
For $\varphi \in L^2(\Rd)$, we put $u= \Fc \varphi$. We decompose 
\begin{equation*}
\varphi= \iFc  Q_{1/h} u +\iFc (u- Q_{1/h} u).
\end{equation*}
We then see that
\begin{equation}
\begin{split}
\Vert  \mathcal F_h \varphi - \mathcal F \varphi \Vert_{L^2(\Rd)} 
&\le
\Vert  (\mathcal F_h - \mathcal F) \iFc  Q_{1/h} u  \Vert_{L^2(\Rd)} \\
&\qquad+
\Vert  (\mathcal F_h - \mathcal F) \iFc  (u-Q_{1/h} u ) \Vert_{L^2(\Rd)}.
\end{split}
\end{equation}
Then it is clear that 
(\ref{eq:fhnfx10}) follows from  (\ref{eq:fhnfx11}),  the fact that
\begin{equation}
\Vert (\Fh -\Fc) \iFc \Vert_{\mathbf B(L^2(\Rd), L^2(\Rd)) }\le 2,
\end{equation}
and  Lemma \ref{lem:Ph}. $\blacksquare$

\vspace{10pt}
\section{Resolvent convergences of ${\mathbb D}_{m,h}$ }\label{sec:main0}

The continuum Dirac operator we shall  consider  in this section 
is      
\begin{equation} 
{\mathbb D}_{m} =  -i \sigma_1 \frac{\partial}{\partial x_1} 
-i\sigma_2  \frac{\partial}{\partial x_2} 
+ m \sigma_3
\;\text{  in }
 L^2(\mathbb R^2)^2, 
\end{equation}
where $m\ge 0$  and 
\begin{equation}
\sigma_1 =
\begin{pmatrix}
0&1 \\ 1& 0
\end{pmatrix}, \,\,\,
\sigma_2 =
\begin{pmatrix}
0& -i  \\ i&0
\end{pmatrix}, \,\,\,
\sigma_3 =
\begin{pmatrix}
1&0 \\ 0&-1
\end{pmatrix}.
\end{equation}
It is well-known that ${\mathbb D}_{m}$ is a self-adjoint operator in  $L^2(\mathbb R^2)^2$ 
with domain  $H^1(\mathbb R^2)^2$,
 the Sobolev space of order $1$ of $\mathbb C^2$-valued functions.

The discrete Dirac operator  ${\mathbb D}_{m,h}$   
  we shall consider is   
\begin{equation}\label{eq:d-dirac2}
{\mathbb D}_{m,h} = 
\begin{pmatrix}
m &  i \partial_{1,h}^* + \partial_{2,h}^* \\
- i \partial_{1,h} + \partial_{2,h}  & -m
\end{pmatrix}
\text{  in }
 L^2(\Zh^2)^2, 
\end{equation}
where difference operators $\partial_{j,h}$ and $\partial_{j,h}^*$  ($j \in \{ 1, \, 2 \}$) are  as defined in (\ref{eq:diffops}). 
It is evident that ${\mathbb D}_{m,h}$ is a bounded self-adjoint operator in 
 $L^2(\Zh^2)^2$.
We mention in passing that  (\ref{eq:d-dirac2}) is a Dirac operator with supersymmetry
in the abstract sense (see \cite[Chapter 5]{Thaller}).
 It  can be rewritten 
in the form 
\begin{equation*}
{\mathbb D}_{m,h} = 
-i\sigma_1
\begin{pmatrix}
 \partial_{1,h}&  0 \\
0 & - \partial_{1,h}^* 
\end{pmatrix}
-i\sigma_2
\begin{pmatrix}
 \partial_{2,h}&  0 \\
0 & - \partial_{2,h}^* 
\end{pmatrix}
+
m\sigma_3,
\end{equation*}
which is comparable  with (\ref{eq:dirac2}).

In accordance with the decomposition (\ref{eq:embed1}) of $L^2(\Rd)$, 
we can compare $\Dc$ and $\Dh\oplus \mathbf 0_h$
in the same Hilbert space  $L^2(\Rtw)^2$, where $\mathbf 0_h$ is the null operator on $(L^2(\mathbb Z_h^2)^2)^\bot$. 
In particular, we investigate the difference  
\begin{equation}\label{eq:differnc0}
(\Dh-z)^{-1}\oplus \mathbf 0_h -(\Dc -z)^{-1}  \quad \mbox{as } h \to 0. 
\end{equation} 
We define 
$\wDc := \Fc\Dc\iFc$, which is the operator of multiplication with the matrix-valued function
\begin{equation}\label{eq:dirac2-xi}
\wDc(\xi)
=
\begin{pmatrix}
m &   \xi_1 -i \xi_2\\
   \xi_1 + i \xi_2 & -m
\end{pmatrix}  
\quad (\xi\in\Rtw)
\end{equation}
in $L^2(\Rtw)^2$.

  With help of (\ref{eq:dfourier1}), (\ref{eq:dfourier1-1}) and (\ref{eq:d-dirac2}), we also define
$\wDh :=\Fh{\mathbb D}_{m,h} \iFh$,
where we abbreviate $\Fh := \Fh\otimes\mathbf 1_{\Ctw} \in
\mathbf B\big(L^2(\Zhtw)^2, L^2(\Thtw)^2 \big)$ 
and $\iFh := \iFh\otimes\mathbf 1_{\Ctw} \in
\mathbf B\big(L^2(\Thtw)^2, L^2(\Zhtw)^2 \big)$.
The operator $\wDh\in \mathbf B(L^2(\Thtw)^2)$ is the operator of multiplication with the matrix-valued function
\begin{equation}\label{eq:d-dirac2-xi}
\wDh(\xi) =
\begin{pmatrix}
m &   \dfrac{ i(e^{-ih\xi_1} -1 ) +   (e^{-ih\xi_2} -1)}{h}\\
  \dfrac{-i (e^{ih\xi_1} -1 ) +  (e^{ih\xi_2} -1)}{h} & -m
\end{pmatrix}
\end{equation}
$(\xi\in\Thtw)$.
With the notation of  (\ref{eq:dfourierL2}), we see that
\begin{equation}
{\mathbb D}_{m,h} \oplus \mathbf 0_h = \iFh(\wDh \oplus \mathbf 0_h )\Fh  \; \mbox{ in }  L^2(\Rtw)^2. 
\end{equation}
Therefore, the difference in (\ref{eq:differnc0}) can be written as
\begin{equation} \label{eq:differnc1}
\begin{split}
&(\Dh-z)^{-1}\oplus \mathbf 0_h -(\Dc -z)^{-1}  \\
&\quad
=
\iFh\big( (\wDh-z)^{-1} \oplus \mathbf 0_h )\Fh - \iFc (\wDc -z)^{-1}\Fc.
\end{split}
\end{equation}

To investigate the matrices $\wDc(\xi)$ and $\wDh(\xi)$, we  start by noting the following.

\begin{lem} \label{lem:FWT-0}
Let $m\ge 0$ and $\zeta \in \mathbb C$, and assume that either  $m>0$ or $\zeta \not= 0$. 
Then,  with the unitary matrix 
 \begin{equation}\label{eq:FWT-1}
 U_m(\zeta) = 
 \frac{1}{\sqrt{2\,}\sqrt{\mu_m(\zeta)^2+m \mu_m(\zeta) }}
 \begin{pmatrix}
\mu_m(\zeta) +m  &    - \overline{\zeta}  \\
\zeta  &  \mu_m(\zeta) +m 
\end{pmatrix},
 \end{equation}
where $\mu_m(\zeta)=\sqrt{|\zeta|^2+ m^2 }$, we have 
\begin{equation}
U_m(\zeta)^* \!
\begin{pmatrix}
m & \overline{\zeta}  \,\\
\zeta & -m \,
\end{pmatrix}
U_m(\zeta)=
\begin{pmatrix}
\mu_m(\zeta) & 0 \,\\
0 & - \mu_m(\zeta)\,
\end{pmatrix}.
\end{equation}
\end{lem}

\vspace{10pt}
Note that the columns of the matrix on the right hand side of (\ref{eq:FWT-1}) are the eigenvectors of the matrix
\begin{equation}
\begin{pmatrix}
m &  \overline{\zeta}  \,\\
\zeta & -m \,
\end{pmatrix}.
\end{equation}

Applying Lemma \ref{lem:FWT-0} to the matrix  $\wDc(\xi)$ in (\ref{eq:dirac2-xi}) with $\zeta=\xi_1 + i\xi_2$, we see that
the eigenvalues of the matrix  $\wDc(\xi)$ are given by 
\begin{equation}
\pm\lambda_m(\xi):=\pm\sqrt{|\xi|^2+ m^2 \,},
\end{equation}
and that the unitary transformation defined by 
 $U_m(\xi_1 + i \xi_2)^* [\Fc \varphi](\xi)$ $(\varphi \in L^2(\Rtw)^2)$ brings 
 the Dirac operator $\Dc$ in (\ref{eq:dirac2}) into the form of
 the operator of multiplication with the diagonal matrix-valued function
\begin{equation}\label{eq:diag-c}
\begin{pmatrix}
\lambda_m(\xi)  & 0 \,\\
0  & -\lambda_m(\xi)  \,
\end{pmatrix}
\end{equation}
in $L^2(\Rtw)^2$.
 This unitary transformation  is a two-dimensional
 version of the Foldy-Wouthuysen-Tani transformation; 
 see  \cite{Foldy},   \cite[Section 1.4]{Thaller},  \cite[Section2.1]{Ben-Um}.
 As is well-known,
one can infer from (\ref{eq:diag-c}) that the Dirac operator $\Dc$ is absolutely continuous and  that its 
spectrum is given by
$\sigma(\Dc)=(-\infty, \, -m] \cup [m, \, \infty)$.

In a similar manner,
applying Lemma \ref{lem:FWT-0} to the matrix  $\wDh(\xi)$ in (\ref{eq:d-dirac2-xi}) 
with 
\begin{equation} \label{eq:d-dirac2-xi0}
\zeta= \dfrac{-i (e^{ih\xi_1} -1 ) +  (e^{ih\xi_2} -1)}{h},
\end{equation}
we see that
the eigenvalues of the matrix  $\wDh(\xi)$ are given by  
\begin{equation}
\pm\lambda_{m,h}(\xi):=\pm\sqrt{h^{-2}\omega(h\xi)+m^2},
\end{equation}
where    
\begin{equation}\label{eq:omega-0}
\begin{split}
\omega(\xi) 
&= 
4 +2 \sin(\xi_1 -\xi_2) - 2 (\sin \xi_1 + \cos \xi_1) + 2 (\sin \xi_2 - \cos \xi_2) \\
&=2(1 -\cos \xi_1)(1+\sin \xi_2) + 2 (1-\cos \xi_2) (1 - \sin \xi_1).
\end{split}
\end{equation}
Then the unitary transformation defined by 
 $U_m(\zeta)^* [\Fh f](\xi)$ ${(f \in L^2(\Zhtw)^2)}$, with $\zeta$ as specified in \ref{eq:d-dirac2-xi0}, brings 
 the discrete Dirac operator $\Dh$ in (\ref{eq:d-dirac2}) into the form of
 the operator of multiplication with the diagonal matrix-valued function 
\begin{equation}\label{eq:diag-d}
\begin{pmatrix}
\lambda_{m,h}(\xi)  & 0 \,\\
0  & -\lambda_{m,h}(\xi)  \,
\end{pmatrix}
\end{equation}
 in $L^2(\Thtw)^2$.

\begin{rem}
It is interesting that the Laplacian on the hexagonal lattice considered in {\rm \cite{IJ}} has a 
similar form to the discrete Dirac operator (\ref{eq:d-dirac2}) in the 
massless case $m=0$. As a result, a function similar to $\omega(\xi)$ appears in  
{\rm \cite[subsection 8.1]{IJ}}. 
\end{rem}

Simple calculations show that  for $\xi \in \Ttw_1$
\begin{equation*}
\begin{split}
&\nabla \omega(\xi) = 0  \Longleftrightarrow  \\
&\quad \xi\in\left\{(0,\,0),  \,  \Big( \frac{\pi}{2}, \,- \frac{\pi}{2} \Big), \,
 \Big( \! - \! \dfrac{3\pi}{4}, \,\dfrac{3\pi}{4} \Big),\, 
\Big( \dfrac{\pi}{4}, \,-\dfrac{\pi}{4} \Big),
\, \Big( \dfrac{\pi}{4}, \,\dfrac{3\pi}{4} \Big),   \,  \Big( \!- \! \dfrac{3\pi}{4}, \, -\dfrac{\pi}{4} \Big) \right\},
\end{split}
\end{equation*}
and that   the function
$\omega : \Ttw_1 \rightarrow \mathbb R$  has the following properties:
\begin{list}{}{ }
\item[(1)] $\omega$ attains its  minimum value $0$ at $(0,\,0)$ and at 
 $\Big( \dfrac{\pi}{2}, \,- \dfrac{\pi}{2}\Big)$;
\vspace{4pt}
\item[(2)] $\omega$ attains its unique maximum  $6 + 4\sqrt{2}$ at $\Big(\! - \!\dfrac{3\pi}{4}, \,\dfrac{3\pi}{4}\Big)$;
\vspace{4pt}
\item[(3)] The saddle points of $\omega$ are $\Big( \dfrac{\pi}{4}, \,-\dfrac{\pi}{4}\Big), \,
\Big( \dfrac{\pi}{4}, \,\dfrac{3\pi}{4}\Big),   \, 
\Big( \!- \!\dfrac{3\pi}{4}, \, -\dfrac{\pi}{4}\Big) $.
\end{list}

\vspace{15pt}
Summing up, we have shown  the following.  

\begin{thm} \label{thm:Dmh}
The discrete Dirac operator  ${\mathbb D}_{m,h}$  is a bounded  self-adjoint operator in $L^2(\Zh^2)^2$ 
with
purely absolutely continuous spectrum
$$\sigma({\mathbb D}_{m,h})=\Big[\!  - \! \sqrt{h^{-2}(6+4\sqrt{2}) +m^2}, \,-m \Big]
\cup 
\Big[m, \,  \! \sqrt{h^{-2}(6+4\sqrt{2}) +m^2}  \,\Big].$$
\end{thm}

\begin{rem}
As was mentioned in the introduction,  some spectral properties of 
2D and 3D discrete Dirac operators 
with a fixed mesh size were already discussed in \cite[Theorem 2.1]{POC},  where
the statement of Theorem \ref{thm:Dmh} is shown in the special case  $h=1$. 
\end{rem}

To prove the convergence theorems on ${\mathbb D}_{m,h}$, 
we need to examine the function  $\omega$ in more detail. 
It follows from (\ref{eq:omega-0}) that
\begin{equation}\label{eq:omega-1}
0\le\omega(\xi) \le 2 |\xi|^2    \quad  \;\; (\xi \in \Rtw) 
\end{equation}
and that
\begin{equation}\label{eq:omega-2}
\omega(\xi) \ge \frac{2-\sqrt{2}}{8} |\xi|^2      \quad\;\; (|\xi|_{\infty}\le\frac{\pi}{4} ) 
\end{equation}
 (Obviously, the number $(2-\sqrt{2})/8$ can be replaced by any smaller positive constant, but we fix this value for the sake of definiteness.) 
Also, it follows from (\ref{eq:omega-0})  that, setting $\alpha = \left(\frac\pi 2, -\frac\pi 2\right)$,
\begin{equation}
\begin{split}
\omega(\alpha + \xi )
&= 
4 -2 \sin(\xi_1 -\xi_2) + 2 (\sin \xi_1 - \cos \xi_1) - 2 (\sin \xi_2 + \cos \xi_2) \\
&=
2(1 -\cos \xi_2)(1+\sin \xi_1) + 2 (1-\cos \xi_1) (1 - \sin \xi_2),
\end{split}
\end{equation}
which, together with (\ref{eq:omega-1}),  implies that
\begin{equation}\label{eq:omega-3}
0\le\omega (\alpha + \xi ) 
\le 2 |\xi|^2    \quad  \;\; (\xi \in \Rtw) 
\end{equation}
and that
\begin{equation}\label{eq:omega-4}
\omega (\alpha + \xi ) 
 \ge \frac{2-\sqrt{2}}{8} |\xi|^2      \quad\;\; (|\xi|_{\infty}\le\frac{\pi}{4} ).
\end{equation}

Now, for any $\varepsilon \in (0, \pi^2/128)$
we divide  $\Ttw_1$ into two disjoint subsets, 
\begin{equation}\label{eq:subset0}
\begin{split}
&\hspace{80pt}
 \Ttw_1 
=  E(\varepsilon) \cup F(\varepsilon),  \\
&
E(\varepsilon):= 
\big\{ \, \xi \in  \Ttw_1  \, \big| \,  \omega (\xi)  \ge \varepsilon \, \big\},   \;\;
F(\varepsilon):= \big\{ \, \xi \in  \Ttw_1  \, \big| \,  \omega (\xi)  < \varepsilon \, \big\}.
\end{split}
\end{equation}
In view of (\ref{eq:omega-1}), (\ref{eq:omega-2}),(\ref{eq:omega-3}) and (\ref{eq:omega-4}), 
one can see that $F(\varepsilon)$ consists of two disjoint components 
$F_0(\varepsilon)$ and $F_1(\varepsilon)$ 
satisfying   
\begin{equation} \label{eq:subset1}
B\Big( 0, \, \sqrt{\dfrac{\varepsilon}{2}} \, \Big)  \subset F_0(\varepsilon)  \subset  B(0, \, 4\sqrt{\varepsilon}),
\end{equation}
and 
\begin{equation} \label{eq:subset2}
B\Big(\alpha, \, \sqrt{\dfrac{\varepsilon}{2}} \, \Big) \subset F_1(\varepsilon)  \subset  B(\alpha, \, 4\sqrt{\varepsilon}), 
\end{equation}
where $B(a, \, r)= \big\{ \, \xi \in \Rtw \, \big| \,  | a - \xi |  < r \, \big\}$  is the ball 
 with center $a$ and radius $r>0$.
Hence we have a disjoint decomposition of $\Ttw_1 $:
\begin{equation}
\Ttw_1  = E(\varepsilon)\cup F_0(\varepsilon) \cup F_1(\varepsilon).
\end{equation}
Accordingly, we have 
a disjoint decomposition of $\Thtw$:
\begin{equation}\label{eq:disjoint0}
\Thtw = E(\varepsilon, 1/h)\cup F_0(\varepsilon, 1/h) \cup F_1(\varepsilon, 1/h),
\end{equation}
where
\begin{equation}
\begin{split}
&
E(\varepsilon, 1/h):=
\big\{ \, \xi \in \Thtw \, \big| \, h\xi \in E(\varepsilon) \, \big\},   \\
&
F_j(\varepsilon, 1/h):=
\big\{ \, \xi \in \Thtw \, \big| \, h\xi \in F_j(\varepsilon) \, \big\}  \qquad  (j=0,\,1).
\end{split}
\end{equation}

\vspace{10pt}
 In the following proposition, we refer to the decomposition $L^2(\Rtw)^2 = L^2(\Thtw)^2 \oplus L^2(\Rtw\setminus\Thtw)^2$; remember that $Q_{1/h}$, the operator of multiplication with $\chi_{\Thtw}$, is the orthogonal projection
onto the first direct summand. 
\begin{prop} \label{prp:xi}
Let $z \in \mathbb C \setminus \mathbb R$. Then, for any $u \in  L^2(\Rtw)^2$,
\begin{equation}
\lim_{h\to 0}
\Vert 
\big[ (\widehat{\mathbb D}_{m,h} - z )^{-1} \!\oplus  \mathbf 0_h -  (\widehat{\mathbb D}_{m} - z )^{-1}\big] u 
 \, \Vert_{L^2(\Rtw)^2}=0.
\end{equation}
\end{prop}

\vspace{3pt}

The proof of Proposition \ref{prp:xi} can be found after the proof of Lemma \ref{lem:differ-xi3}.

\vspace{1pt}

\begin{lem} \label{lem:differ-xi1}
Let $h>0$. Then we have
\begin{equation}
\big\Vert \wDh (\xi) - \wDc (\xi)  \big\Vert_{\mathbf B(\Ctw)} 
\le
\dfrac{\, h}{2} |\xi|^2   \qquad (\xi \in \Thtw),
\end{equation}
where  $\Vert M \Vert_{\mathbf B(\Ctw)}$ denotes the operator norm of a $2\times 2$ matrix $M$ as
a linear operator in $\Ctw$.
\end{lem}

\noindent
{\bf Proof.} 
By using the inequality
\begin{equation} \label{eq:expo-0}
| e^{i\theta} -1 -i\theta | \le \frac{\,\theta^2}{2}  \qquad ( \theta \in \mathbb R),
\end{equation}
we see that
\begin{equation} \label{eq:diracs-differ0}
\begin{split}
&
\Big|
\dfrac{-i (e^{ih\xi_1} -1 ) +    (e^{ih\xi_2} - 1) }{h}  - (\xi_1 + i \xi_2)
\Big|  \\
&\quad \le
\frac{1}{h} \Big\{
\frac{\, (h\xi_1)^2}{2} + \frac{\, (h\xi_2)^2}{2}
\Big\}
= 
\frac{\, h}{2} |\xi|^2  \qquad (\xi \in \Rtw),
\end{split}
\end{equation}
which, together with (\ref{eq:dirac2-xi}) and (\ref{eq:d-dirac2-xi}),
implies the lemma. 
$\blacksquare$

\medskip
 Since  $h |\xi| \le \sqrt{2}\pi$ for $\xi \in \Thtw$, the above lemma immediately gives the following estimate.

\begin{lem}\label{lem:differ-xi1+}
Let $h>0$. Then we have
\begin{equation}
\big\Vert \wDh (\xi) - \wDc (\xi)  \big\Vert_{\mathbf B(\Ctw)} 
\le
\dfrac{\, \sqrt{2}\pi}{2} |\xi|   \qquad (\xi \in \Thtw).
\end{equation}
\end{lem}

\vspace{5pt}
\begin{lem}\label{lem:differ-xi2}
For any $z \in \mathbb C \setminus \mathbb R$, there exists a constant  $C_z$ such that 
 \begin{equation}\label{ineq:Dm-xi-00}
 \Vert (\wDc (\xi) - z)^{-1}\Vert_{\mathbf B(\Ctw)}
 \le 
 C_z  \big( |\xi|^2 + m^2 + | \mathfrak{Im}\,z|^2 \big)^{-1/2}
 \qquad
(\xi \in  \Rtw).
 \end{equation}
\end{lem}

\noindent
 {\bf Proof.}
 We first note that the matrix $\wDc (\xi) - z$ is unitarily equivalent to the matrix
\begin{equation}\label{eq:diag-d-z}
\begin{pmatrix}
\lambda_{m}(\xi)  -z & 0 \,\\
0  & -\lambda_{m}(\xi) -z  \,
\end{pmatrix},
\end{equation}
 as was discussed after Lemma \ref{lem:FWT-0}.
Hence it is clear that
\begin{equation}\label{eq:resnm}
\|(\wDc(\xi) - z)^{-1}\|_{\mathbf B(\Ctw)} = \max\left\{\frac 1 {|\lambda_m(\xi) - z|}, \frac 1 {|-\lambda_m(\xi) - z|}\right\}.
\end{equation}
Now let $z_0 = i \mathop{\mathfrak{Im}} z$. Then
\begin{equation*}
\frac{|z_0 - t|^2}{|z - t|^2} = \frac{(\mathop{\mathfrak{Im}} z)^2 + t^2}{(\mathop{\mathfrak{Im}} z)^2 + (t - \mathop{\mathfrak{Re}} z)^2}
\end{equation*}
is a continuous function of $t\in\mathbb R$ that tends to $1$ as $t \to \pm\infty$.
It is therefore bounded, so there exists a constant $C_z$ such that
$|z_0 - t| \le C_z\,|z - t|$ for all $t\in\mathbb R$, which together with (\ref{eq:resnm}) gives (\ref{ineq:Dm-xi-00}).
 $\blacksquare$

  \vspace{10pt}
 \begin{lem}\label{lem:differ-xi3}
Let $z \in \mathbb C \setminus \mathbb R$,  and let  $\varepsilon>0$. Then
for any $h \in \Big(0, \,  \dfrac{\sqrt{\varepsilon}}{2| \mathfrak{Re}\,z|} \Big)$ we have
 \begin{equation}\label{ineq:Dmh-0}
 \Vert (\wDh (\xi) - z)^{-1}\Vert_{\mathbf B(\Ctw)}
 \le
 \min \Big(\frac{1}{|\mathfrak{Im}\,z |}, \, \frac{\,2 h \,}{\sqrt{\varepsilon} \,} \Big)
 \qquad
\big(\xi \in  E(\varepsilon, 1/h) \big).
\end{equation}
 \end{lem}

 \noindent
 {\bf Proof.}
As a consequence of the spectral theorem for self-adjoint operators, 
 \begin{equation}\label{ineq:Dmh-0-1}
\Vert (\wDh (\xi) - z)^{-1}\Vert_{\mathbf B(\Ctw)}\le \frac{1}{|\mathfrak{Im}\,z |}
\qquad (\xi \in \Thtw)  
\end{equation}
 for any   $z \in \mathbb C \setminus \mathbb R$. 
 Therefore, it is sufficient to prove the inequality
 \begin{equation}\label{ineq:Dmh-0-2}
  \Vert (\wDh (\xi) - z)^{-1}\Vert_{\mathbf B(\Ctw)}
 \le
 \frac{\,2h \,}{\sqrt{\varepsilon} \,} 
 \qquad  
 (\xi \in  E(\varepsilon, 1/h)).
 \end{equation}
To this end,  
 we  note  the fact that the matrix $\wDh (\xi) - z$ is unitarily equivalent to the matrix
\begin{equation}\label{eq:diag-d-z0}
\begin{pmatrix}
\lambda_{m,h}(\xi)  -z & 0 \,\\
0  & -\lambda_{m,h}(\xi) -z  \,
\end{pmatrix},
\end{equation}
 as was discussed after Lemma \ref{lem:FWT-0}.
 By the reverse triangle inequality, 
 \begin{equation}\label{ineq:Dmh-1}
 \begin{split}
|\lambda_{m,h}(\xi)  \pm z | 
& >   \sqrt{h^{-2}\,\omega(h\xi) + m^2 \,} - | \mathfrak{Re}\,z |  \\
 & \quad 
 >  h^{-1}\sqrt{\varepsilon} - | \mathfrak{Re}\,z |   \qquad ( \mbox{for }  \xi \in E(\varepsilon, 1/h) ).
 \end{split}
 \end{equation}
If $h$ satisfies the inequality $0 < h < \sqrt{\varepsilon}/(2| \mathfrak{Re}\,z| )$,
then 
 the inequality (\ref{ineq:Dmh-0-2}) follows from (\ref{ineq:Dmh-1}).
 $\blacksquare$

\vspace{20pt}
\noindent 
{\bf Proof of Proposition \ref{prp:xi}.} 
 Let $u\in L^2(\Rtw)^2$. Then
\begin{align*}
&\left\|\left((\wDh - z)^{-1}\oplus \mathbf 0_h - (\wDc - z)^{-1}\right) u\right\|_{L^2(\Rtw)^2}^2 \\
 &\qquad= \int_{\Thtw} \left|\left( (\wDh(\xi) - z)^{-1} - (\wDc(\xi) - z)^{-1}\right) 
 u(\xi)\right|_{\Ctw}^2 d\xi \\
 &\qquad\qquad+ \int_{\Rtw\setminus\Thtw} \left|(\wDc(\xi) - z)^{-1}\,u(\xi)\right|_{\Ctw}^2 d\xi.
\end{align*}
As the operator norm of $(\wDc(\xi) - z)^{-1}$ can be estimated by $|\mathop{\mathfrak{Im}} z|^{-1}$, the second
integral tends to $0$ as $h\to 0$.
Let $\varepsilon \in (0, \pi^2/128)$.
In view of (\ref{eq:disjoint0}), we divide the first integral into three terms,
 \begin{align}\label{eq:conv-xi-sp-3}
&\Big(\int_{E(\varepsilon, 1/h)}  + \int_{F_0(\varepsilon, 1/h)}  +  \int_{F_1(\varepsilon, 1/h)}\Big)
\big|
 \{( \wDh(\xi) - z )^{-1} 
-  
 ( \wDc(\xi) - z )^{-1} \}
 u(\xi)  \big|_{{\Ctw}}^2   \,  d\xi    \nonumber \\
 & 
 \qquad =:
 I(\varepsilon, h) +  I\!I_0(\varepsilon, h) +  I\!I_1(\varepsilon, h).
 \end{align}

Since
\begin{equation}
\begin{split}
&(\wDh (\xi) - z)^{-1} -  (\wDc (\xi) - z)^{-1} \\
&\quad = 
\big( \wDh (\xi) - z)^{-1} ( \wDc (\xi) - \wDh (\xi)  \big) 
(\wDc (\xi) - z)^{-1},
\end{split}
\end{equation}
we get
\begin{equation} \label{eq:conv-I-eh1}
\begin{split}
 I(\varepsilon, h) 
 &\le  
\int_{E(\varepsilon, 1/h)} 
\Vert \big( \wDh (\xi) - z)^{-1}\Vert_{\mathbf B(\Ctw)}^2  \\
&
\quad \times
 \Vert  \wDc (\xi) - \wDh (\xi)  \Vert_{\mathbf B(\Ctw)}^2 \, 
\Vert \wDc (\xi) - z)^{-1}  \Vert_{\mathbf B(\Ctw)}^2 \,
 \big|u(\xi)  \big|_{{\Ctw}}^2   \,  d\xi    \\
  &\le  
\int_{E(\varepsilon, 1/h)} 
\Big\{ \min \Big(\frac{1}{|\mathfrak{Im}\,z |}, \, \frac{\,2 h \,}{\sqrt{\varepsilon} \,} \Big) \Big\}^2  \\
 &
\qquad \times
\Big( \dfrac{\, \sqrt{2}\pi}{2} |\xi|  \Big)^2
C_z^2  \big( |\xi|^2 + m^2 + | \mathfrak{Im}\,z|^2 \big)^{-1}
 \big|u(\xi)  \big|_{{\Ctw}}^2   \,  d\xi   \\
  &\le  
 \Big\{ \min \Big(\frac{1}{|\mathfrak{Im}\,z |}, \, \frac{\,2 h \,}{\sqrt{\varepsilon} \,} \Big) \Big\}^2 
  \Big( \dfrac{\, \sqrt{2}\pi}{2}  \Big)^2  
  C_z^2 
\int_{\Rtw} \big|u(\xi)  \big|_{{\Ctw}}^2   \,  d\xi ,
\end{split}
\end{equation}
where we have used Lemmas \ref{lem:differ-xi1+},  \ref{lem:differ-xi2} 
 (with the same constant $C_z$) and 
 \ref{lem:differ-xi3}. 
It follows from (\ref{eq:conv-I-eh1})
that
\begin{equation}\label{eq:conv-I-eh2}
 \lim_{h\to 0} I(\varepsilon, h)=0.
 \end{equation}
The second term in (\ref{eq:conv-xi-sp-3}) can be estimated in a similar manner:  
\begin{equation} \label{eq:conv-II0-eh1}
\begin{split}
I\!I_0(\varepsilon, h)
 &\le  
\int_{F_0(\varepsilon, 1/h)} 
\Vert \big( \wDh (\xi) - z)^{-1}\Vert_{\mathbf B(\Ctw)}^2  \\
&
\quad \times
 \Vert  \wDc (\xi) - \wDh (\xi)  \Vert_{\mathbf B(\Ctw)}^2 \, 
\Vert \wDc (\xi) - z)^{-1}  \Vert_{\mathbf B(\Ctw)}^2 \,
 \big|u(\xi)  \big|_{{\Ctw}}^2   \,  d\xi    \\
  &\le  
\int_{F_0(\varepsilon, 1/h)} 
  \Big(\frac{1}{|\mathfrak{Im}\,z |} \Big)^2   \\
 &
\qquad  \times
 \Big( \dfrac{h}{2} |\xi|^2  \Big)^2C_z^2  \big( |\xi|^2 + m^2 + | \mathfrak{Im}\,z|^2 \big)^{-1}
 \big|u(\xi)  \big|_{{\Ctw}}^2   \,  d\xi ,
\end{split}
\end{equation}
where we have used  (\ref{ineq:Dmh-0-1}),  Lemmas \ref{lem:differ-xi1} and  \ref{lem:differ-xi2}.
If $\xi \in F_0(\varepsilon, 1/h)$, then by (\ref{eq:subset1}) we see that $h |\xi| < 4\sqrt{\varepsilon}$,  hence
that
\begin{equation}\label{eq:conv-II0-eh2}
\Big( \dfrac{h}{2} |\xi|^2  \Big)^2C_z^2  \big( |\xi|^2 + m^2 + | \mathfrak{Im}\,z|^2 \big)^{-1}
<
4\varepsilon  C_z^2 . 
\end{equation}
It follows from  (\ref{eq:conv-I-eh2}) and (\ref{eq:conv-II0-eh2}) that
\begin{equation}\label{eq:conv-II0-eh3}
 \limsup_{h\to 0} I\!I_0(\varepsilon, h) 
\le 
\Big(\frac{2C_z}{|\mathfrak{Im}\,z |}  \, \Vert u \Vert_{L^2_{\xi}}\Big)^{\!2}  
 \,  \varepsilon. 
\end{equation}

 Also,  the third term in (\ref{eq:conv-xi-sp-3}) can be estimated in a similar manner:  
\begin{equation} \label{eq:conv-II1-eh1}
\begin{split}
I\!I_1(\varepsilon, h)
 &\le  
\int_{F_1(\varepsilon, 1/h)} 
\Vert \big( \wDh (\xi) - z)^{-1}\Vert_{\mathbf B(\Ctw)}^2  \\
&
\quad \times
 \Vert  \wDc (\xi) - \wDh (\xi)  \Vert_{\mathbf B(\Ctw)}^2 \, 
\Vert \wDc (\xi) - z)^{-1}  \Vert_{\mathbf B(\Ctw)}^2 \,
 \big|u(\xi)  \big|_{{\Ctw}}^2   \,  d\xi    \\
  &\le  
\int_{F_1(\varepsilon, 1/h)} 
  \Big(\frac{1}{|\mathfrak{Im}\,z |} \Big)^2   \\
 &
\qquad  \times
 \Big( \dfrac{\sqrt{2}\pi}{2} |\xi|  \Big)^2C_z^2  \big( |\xi|^2 + m^2 + | \mathfrak{Im}\,z|^2 \big)^{-1}
 \big|u(\xi)  \big|_{{\Ctw}}^2   \,  d\xi   \\
  &\le  
   \Big(\frac{\sqrt{2}\pi C_z}{2|\mathfrak{Im}\,z |} \Big)^2
\int_{F_1(\varepsilon, 1/h)} 
 \big|u(\xi)  \big|_{{\Ctw}}^2   \,  d\xi  
\end{split}
\end{equation}
where we have used  (\ref{ineq:Dmh-0-1}),  Lemmas \ref{lem:differ-xi1+} and  \ref{lem:differ-xi2}.
Since,  by (\ref{eq:subset2}),
\begin{equation}\label{eq:conv-II1-eh10}
  F_1(\varepsilon, 1/h) 
  \subset  
  \big\{  \, \xi \in \Rtw  \, \big|  \,   |\xi| \ge h^{-1} (|\alpha | - 4 \sqrt{\varepsilon})  \, \big\},
\end{equation}
we get
\begin{equation}\label{eq:conv-II1-eh2}
\lim_{h\to 0} I\!I_1(\varepsilon, h) =  0. 
\end{equation}

We can deduce from (\ref{eq:conv-xi-sp-3}),  
 together with (\ref{eq:conv-I-eh2}),  (\ref{eq:conv-II0-eh3})
and (\ref{eq:conv-II1-eh2}),  that
\begin{equation}
\begin{split}
&\limsup_{h\to 0}\int_{\Thtw} \! \big|
 \{( \wDh(\xi) - z )^{-1} 
-  
 ( \wDc(\xi) - z )^{-1} \}
 u(\xi)  \big|_{{\Ctw}}^2   \,  d\xi \\
& \hspace{90pt}\le
 \Big(\frac{2C_z}{|\mathfrak{Im}\,z |}  \, \Vert u \Vert_{L^2_{\xi}}\Big)^{\!2}  
 \, \varepsilon .
 \end{split}
\end{equation}
 This completes the proof of conclusion  
 of Proposition \ref{prp:xi}, since  $\varepsilon >0$ was arbitrarily small. 
$\blacksquare$


\vspace{6pt}

 \vspace{6pt}

\begin{thm}\label{thm:x} 
Let $z \in \mathbb C \setminus \mathbb R$. Then  
\begin{equation}
 \mathop{\hbox{\rm s-lim}}_{h\to 0} 
\big\{ ({\mathbb D}_{m,h} - z )^{-1}\!\oplus  \mathbf 0_h \big\}
 = 
({\mathbb D}_{m} - z )^{-1}
\; \mbox{ in  } \,  L^2(\mathbb R^2)^2 .
\end{equation}
\end{thm}

 \noindent
 {\bf Proof.}
 Let  $\varphi \in  L^2(\mathbb R^2)^2$.
 It follows from (\ref{eq:differnc1}) that
\begin{align}
&\big\{
(\Dh-z)^{-1}\oplus \mathbf 0_h 
- (\Dc -z)^{-1}  
\big\} \varphi  \nonumber\\
&\quad
=
\iFh\big\{
(\wDh-z)^{-1} \oplus \mathbf 0_h  
\big\}\Fh \varphi
- \iFc (\wDc -z)^{-1}\Fc \varphi    \nonumber \\
&\quad
=
\iFh\big\{
 (\wDh-z)^{-1} \oplus \mathbf 0_h 
\big\} (\Fh - \Fc ) \varphi \label{eq:differnc2}\\
&\qquad\quad
+
(\iFh - \iFc )\big\{
 (\wDh-z)^{-1} \oplus \mathbf 0_h 
\big\}\Fc \varphi   \label{eq:differnc3}\\
&\qquad\qquad
+
\iFc \big\{
 (\wDh-z)^{-1} \oplus \mathbf 0_h  
- 
(\wDc -z)^{-1}
\big\}\Fc \varphi.   \label{eq:differnc4}
\end{align}

The $L^2$ norm of the term in (\ref{eq:differnc2})  can  be estimated by
\begin{align}
&
\Vert
 (\wDh-z)^{-1} \oplus \mathbf 0_h 
\Vert_{\mathbf B(L^2(\Rtw)^2)}
\Vert(\Fh - \Fc)  \varphi \Vert_{L^2(\Rtw)^2}\\
&\le
 \Big(\frac{1}{|\mathfrak{Im}\,z |}\Big) \Vert(\Fh - \Fc)  \varphi \Vert_{L^2(\Rtw)^2}
\end{align}
which, by Lemma  \ref{lem:Fourier},  tends to $0$ as $h\to 0$.

The term in (\ref{eq:differnc3})  can be written as
\begin{align}
&
(\iFh - \iFc )\big\{
 (\wDh-z)^{-1} \oplus \mathbf 0_h 
\big\}\Fc  \varphi \\
&=
(\iFh - \iFc )\big\{
 (\wDc-z)^{-1}  \oplus \mathbf 0_h 
\big\}\Fc  \varphi    \label{eq:differnc5}\\
&\quad +
(\iFh - \iFc )\big\{
 (\wDh-z)^{-1} \oplus \mathbf 0_h  - (\wDc-z)^{-1} 
 \big\}\Fc \varphi,   \label{eq:differnc6}
\end{align}
where, by Lemma \ref{lem:iFourier}, the $L^2$ norm of  the term (\ref{eq:differnc5})  tends 
 to $0$ as $h \to 0$,
and 
the $L^2$ norm of  the term (\ref{eq:differnc6}) is bounded by
\begin{equation} \label{eq:differnc7}
2 \Vert 
\big\{
 (\wDh-z)^{-1} \oplus \mathbf 0_h  - (\wDc-z)^{-1}  
\big\}\Fc \varphi
\Vert_{ L^2(\Rtw)^2 },
\end{equation}
because $\Vert \iFh \Vert_{\mathbf B(L^2(\Rtw)^2, L^2(\Rtw)^2) }=1$.
Proposition \ref{prp:xi}  implies that $L^2$ norm of the term in  (\ref{eq:differnc7})  tends  to $0$ as $h\to 0$.
Therefore, the $L^2$ norm of the term in (\ref{eq:differnc3}) 
 tends to $0$ as $h\to 0$. 

Finally,
Proposition \ref{prp:xi} immediately implies that $L^2$ norm of the term in (\ref{eq:differnc4}) tends to $0$  as $h\to 0$.
$\blacksquare$

\vspace{6pt}
\noindent
\begin{rem}
In view of the strong convergence of the orthogonal projectors $P_h$ (see Lemma \ref{lem:Ph}), the strong limits of
$(\mathbb D_{m,h} \oplus \mathbf 0_h - z)^{-1}$ and of $(\mathbb D_{m,h} - z)^{-1} \oplus \mathbf 0_h$ are the same.
\end{rem}
\vspace{6pt}
We conclude this section by proving that $(\Dh \oplus \mathbf 0_h - z)^{-1}$ does not converge in the operator norm
sense to $(\Dc -z)^{-1}$ as $h \rightarrow 0$. 
We would like to mention that the proof of Theorem \ref{thm:nonoco} below is based on the idea demonstrated in \cite{Yamada} and \cite{Umeda}.

\vspace{6pt}
\begin{thm}\label{thm:nonoco}
Let $z \in \mathbb C \setminus \mathbb R$. Then
\begin{equation} \label{eq:nonoco-0}
\begin{split}
\liminf_{h\to 0}
&\Vert 
({\mathbb D}_{m,h} \oplus \mathbf 0_h - z )^{-1}  
-  ({\mathbb D}_{m} - z)^{-1}
 \, \Vert_{\mathbf B (L^2(\Rtw)^2)} \\
 & \qquad \ge 
 \max \Big( \frac 1 {|m - z|}, \frac 1 {|m + z|} \Big).
 \end{split}
\end{equation}
\end{thm}

\noindent
{\bf Proof.}
For $h > 0$, consider the function
$u_h = \left(\begin{matrix}y_h \\ 0 \end{matrix}\right)\in L^2(\mathbb R^2)^2$, where
\begin{equation}
y_h(x) = \sqrt h\,e^{i\frac\pi{2 h}(x_1 - x_2)}\,e^{-h(x_1^2 + x_2^2)} \qquad (x \in \mathbb R^2).
\end{equation}
Then $\| u_h \|_{L^2(\mathbb R^2)^2} = \sqrt{\frac \pi 2}$ and
\begin{equation}
(\Fc y_h)(\xi) = \frac 1{2\sqrt h}\,e^{-\frac 1{4h}[(\xi_1 - \frac \pi{2h})^2 + (\xi_2 + \frac \pi{2h})^2]}
\qquad (\xi \in \mathbb R^2).
\end{equation}
Further, by (\ref{eq:dirac2-xi}) we find that
\begin{equation}
(\mathbb D_m - z)^{-1} u_h 
= \iFc \frac 1{m^2 -z^2 + \xi_1^2 + \xi_2^2}
 \left(\begin{matrix} m+z& \xi_1 - i \xi_2 \\ \xi_1 + i \xi_2 & -m+ z \end{matrix}\right) \, \left(\begin{matrix} \Fc y_h \\ 0 \end{matrix}\right),
\end{equation}
and, as the Fourier transform is an isometry on $L^2(\mathbb R^2)$, we conclude that
\begin{equation} \label{eq:nonoco-1}
\begin{split}
\|(\mathbb D_m &- z)^{-1} u_h\|_{L^2(\mathbb R^2)^2}^2 \\
&
= \int_{\mathbb R^2} 
\frac {|m +z|^2 + \xi_1^2 + \xi_2^2}{|m^2 - z^2 + \xi_1^2 + \xi_2^2 \,|^2}
\,\frac 1{4h}\,e^{-[(\xi_1 - \frac \pi{2h})^2 + (\xi_2 + \frac \pi{2h})^2]/2h}\,d\xi
\\
&
\le
  \int_{\mathbb R^2}
   \frac{C_z}
   {   |(m^2 - z^2)  + (\eta_1  + \frac \pi {2h})^2 + (\eta_2 - \frac \pi {2h})^2 |}
  \,\frac 1{4h} \,e^{-(\eta_1^2 + \eta_2^2)/2h}\,d\eta. 
\end{split}
\end{equation}
Here we have used the fact that 
\begin{equation*}
\frac{|m +z|^2 + \xi_1^2 + \xi_2^2}{|m^2 - z^2 + \xi_1^2 + \xi_2^2 \,|} \to 1 
\quad \mbox{ as } \xi=(\xi_1, \xi_2) \to \infty,
\end{equation*}
and the fact that $|m^2 - z^2 + \xi_1^2 + \xi_2^2 \,| \ge c_z >0$  for $\forall \xi \in {\mathbb R}^2$.
In fact,  we have
\begin{equation*}
|m^2 - z^2 + \xi_1^2 + \xi_2^2 \,| \ge
\begin{cases}
 |\mathfrak{Im}\,z |^2  & \mbox{if } z \mbox{ is pure imaginary}, \\
 2 | (\mathfrak{Re}\, z ) (\mathfrak{Im}\,z) | & \mbox{if } z \mbox{ is not pure imaginary}.
 \end{cases}
\end{equation*}
The  second integral in (\ref{eq:nonoco-1}) can be written in the form
\begin{equation}\label{eq:nonoco-2}
 \int_{\mathbb R^2}
   \frac{C_z}
   {   |(m^2 - z^2)  + (\sqrt{h} \, \eta_1  + \frac \pi {2h})^2 + (\sqrt{h} \, \eta_2 - \frac \pi {2h})^2 |}
  \,\frac 1{4} \,e^{-(\eta_1^2 + \eta_2^2)/2}\,d\eta.
\end{equation}
The integrand in (\ref{eq:nonoco-2})
tends to 0 pointwise as $h\rightarrow 0$ and is bounded above by the function
$\frac {C_z}{|m^2 -z^2|}\,\frac{1}{4}\,e^{-(\eta_1^2 + \eta_2^2)/2}$, 
so by the dominated convergence theorem
\begin{equation}
\lim_{h\rightarrow 0} \|(\mathbb D_m - z)^{-1} u_h\|_{L^2(\mathbb R^2)^2} = 0.
\end{equation}

Now, in order to apply the discrete Dirac operator, we project $u_h$ into $L^2(\mathbb Z_h^2)^2$.
We see that
$P_h y_h = \sumZtw \widetilde{y}_h (hn) \charnh$, where
\begin{equation*}
\widetilde{y}_h(hn) 
= \sqrt h \left(\int_{n_1}^{n_1+1} e^{i\frac \pi 2 t}\,e^{-h^3 t^2}\,dt\right) 
\!\! \left(\int_{n_2}^{n_2+1} e^{-i\frac \pi 2 t}\,e^{-h^3 t^2}\,dt\right) 
\quad (n = (n_1, \, n_2)\in \mathbb Z^2)
\end{equation*}
by (\ref{eq:embed2}).
Integration by parts 
and the formula 
$$\int_n^{n+1}2at \,e^{-at^2}\,dt=e^{-an^2} - e^{-a(n+1)^2} \qquad (n \in \mathbb Z) $$ 
show
that 
\begin{align*}
&\int_n^{n+1} e^{\pm i \frac \pi 2 t}\,e^{-h^3t^2}\,dt  \qquad (n \in \mathbb Z) 
\\
 &= \pm \frac 2{i\pi}\,\left(e^{\pm i \frac \pi 2 (n+1)}\,e^{-h^3(n+1)^2} - e^{\pm i \frac \pi 2 n}\,e^{-h^3n^2}\right)
\pm \frac 2{i\pi} \int_n^{n+1} e^{\pm i \frac \pi 2 t}\,2 h^3 t\,e^{-h^3t^2}\,dt
\\
 &= \frac 2 \pi\,(1 \pm i)\,e^{\pm i \frac \pi 2 n}\,e^{-h^3 n^2} \pm 
 \frac {4h^3}{i \pi} \int_n^{n+1} \left(e^{\pm i \frac \pi 2 t} - e^{\pm i \frac \pi 2 (n+1)}\right)\,t\,e^{-h^3 t^2}\,dt
\\
&=: \frac 2 \pi\,(1 \pm i)\,e^{\pm i \frac \pi 2 n}\,e^{-h^3 n^2} \pm  \frac {4h^3}{i \pi}
K^{\pm}(n, \,h).
\end{align*}
Hence
\begin{equation} \label{eq:nonoco-3}
P_h u_h = \left(\begin{matrix}\frac 8{\pi^2} (y_h)_h + R_h \\ 0 \end{matrix}\right)
=\frac 8{\pi^2} (u_h)_h  +\left(\begin{matrix} R_h \\ 0 \end{matrix}\right),
\end{equation}
where we use the notation of (\ref{eq:phi_h}) for $(y_h)_h$, $(u_h)_h$   and set 
$R_h := R_{h,a} + R_{h,b} + R_{h,c}$ 
with
\begin{align*}
R_{h,a}(n,h) &= \sqrt h \frac{8 h^3}{\pi^2}\,(i-1)\,e^{i \frac \pi 2 n_1} e^{-h^3 n_1^2} K^-(n_2, h)
\\
R_{h,b}(n,h) &= - \sqrt h \frac{8 h^3}{\pi^2}\,(i+1)\,e^{-i \frac \pi 2 n_2} e^{-h^3 n_2^2} K^+(n_1, h) 
\qquad (n \in \mathbb Z^2)
\\
R_{h,c}(n,h) &=  \sqrt h \frac{16 h^6}{\pi^2}  K^+(n_1, h)  K^-(n_2, h). 
\end{align*}
Using the asymptotics
\begin{equation}
\sum_{k \in \mathbb Z} e^{-\beta k^2} = \sqrt{\frac \pi \beta} + O(1), \quad
\sum_{k=1}^\infty k^2\,e^{-\beta k^2} = \frac{\sqrt \pi}{4 \beta^{3/2}} + O(\beta^{-1}) \quad (\beta\rightarrow 0),
\end{equation}
we find
\begin{equation}
\sum_{n\in\mathbb Z} 
e^{-h^3 n^2} 
= \sqrt{\frac \pi { h^3}} + O(1)
\end{equation}
and
\begin{align*}
\sum_{n\in\mathbb Z} & \big| h^{\frac 3 2}  K^{\pm}(n, h)   \big|^2
\le \sum_{n\in\mathbb Z} \left(2 \int_n^{n+1} h^{\frac 3 2}
 |t| 
\, e^{-(h^{3/2} t)^2 / 2}\,e^{-h^3 t^2/2}\,dt \right)^2
\\
&\le \frac 4 e \sum_{n\in\mathbb Z} \left(\int_n^{n+1} e^{-h^3 t^2/2}\,d t \right)^2 \\
&\le \frac 8 e \sum_{n=0}^\infty e^{-h^3 n^2}
= \frac{4 \sqrt \pi}  e\, h^{-\frac 3 2} + O(1)
\quad (h\rightarrow 0),
\end{align*}
where we have used the inequality $ h^{\frac 3 2} |t| \, e^{-(h^{3/2} t)^2 / 2} \le 1/\sqrt{e}$ in the second inequality.
Hence, bearing in mind (\ref{eq:embed0}),
\begin{equation}
\|R_{h,a}\|_{L^2(\mathbb Z_h^2)}^2, 
\|R_{h,b}\|_{L^2(\mathbb R^2)}^2
\le
h^3\,\frac{256}{\sqrt 2 \pi^3 e} + O(h^{9/2})
\end{equation}
and
$\|R_{h,c}\|_{L^2(\mathbb Z_h^2)}^2 = O(h^{6})$, which gives
$\lim\limits_{h\rightarrow 0} \|R_{h}\|_{L^2(\mathbb Z_h^2)} = 0$, while
\begin{equation}
\left\|\frac 8{\pi^2} (y_h)_h\right\|_{L^2(\mathbb Z_h^2)} = \frac 8 {\sqrt 2 \pi^{3/2}} + o(1) \qquad (h\rightarrow 0).
\end{equation}
Now, applying the discrete Dirac operator to $(u_h)_h = \left(\begin{matrix} (y_h)_h \\ 0 \end{matrix}\right)$, we find
\begin{equation}
(\mathbb D_{m,h} - z) (u_h)_h = (m - z) (u_h)_h + \left(\begin{matrix} 0 \\ (-i\partial_{1,h} + \partial_{2,h}) (y_h)_h \end{matrix}\right),
\end{equation}
where
\begin{align*}
(-i\partial_{1,h} &+ \partial_{2,h}) (y_h)_h (hn )
= (-i\partial_{1,h} + \partial_{2,h}) \sqrt h \frac 8{\pi^2}\,e^{i\frac \pi 2 (n_1 - n_2)} e^{-h^3(n_1^2+n_2^2)}
\\
&= \frac 1{\sqrt h}\,\frac 8{\pi^2}\,e^{i\frac \pi 2 (n_1 - n_2)} e^{-h^3(n_1^2 + n_2^2)} 
\left((e^{-h^3(2n_1 + 1)} - 1) -i(e^{-h^3(2n_2+1)} - 1) \right)
\\
&=\frac 1{\sqrt h}\,\frac 8{\pi^2}\,e^{i\frac \pi 2 (n_1 - n_2)}  
\\ 
&\qquad \times
\left(  (e^{-h^3(n_1+1)^2} - e^{-h^3n_1^2}) e^{-h^3n_2^2} 
 - i  e^{-h^3n_1^2}  (e^{-h^3(n_2+1)^2} - e^{-h^3n_2^2})  \right)
\\
&=: \widetilde{R}_h (n,h) 
\quad (n =(n_1, \,n_2) \in \mathbb Z^2).
\end{align*}
Then, using   the formula
\begin{equation*}
\sum_{n_1\in \mathbb Z}  \big |  e^{-h^3(n_1+1)^2} -  e^{-h^3 n_1^2} \big|^2 
= 2 \sum_{n_1=0}^{\infty} 
\big |  e^{-h^3 (n_1+1)^2} -  e^{-h^3  n_1^2} \big|^2   
\end{equation*}
and
the estimate
\begin{equation}
 \big|  e^{-h^3(2n_1+1)} - 1 \big| 
= \Big| \int_0^{2n_1+1} (-h^3)\,e^{-h^3 s}\,ds \Big| 
\le h^3\,(2n_1+1)   
\qquad (n_1\ge 0), 
\end{equation}
we obtain for
$\widetilde{R}_{h,a}(n,h) := \frac 1{\sqrt h}\,\frac 8{\pi^2}\,
e^{i \frac \pi 2  n_1 }  (e^{-h^3(n_1+1)^2} - e^{-h^3n_1^2}) e^{-h^3n_2^2}$
\begin{align*}
\| \widetilde{R}_{h,a}\|_{L^2(\mathbb Z_h^2)}^2 
&\le h^7\,\frac{64}{\pi^4} \left(
2\sum_{n_1=0}^{\infty} (2n_1+1)^2\,e^{-2h^3 n_1^2} 
\right)
\left(\sum_{n_2\in\mathbb Z} e^{-2h^3 n_2^2}\right)
\\
&\le h^7\,\frac{64}{\pi^4} \left(
2 \sum_{n_1=0}^{\infty} e^{-2h^3 n_1^2} + 16 \sum_{n_1=0}^{\infty} n_1^2\,e^{-2h^3 n_1^2}
\right)\left(\sum_{n_2\in\mathbb Z} e^{-2h^3 n_2^2}\right)
\\
&= h\,\frac{64}{\pi^3} + o(h) \qquad (h\rightarrow 0)
\end{align*}
and similarly for
$\widetilde{R}_{h,b}(n,h) := \frac 1{\sqrt h}\,\frac 8{\pi^2}\,
e^{ -i \frac \pi 2  n_2 } e^{-h^3n_1^2} (e^{-h^3(n_2+1)^2} - e^{-h^3n_2^2}) $,
so
\begin{equation}
\lim_{h\rightarrow 0} \|
\widetilde{R}_h 
\|_{L^2(\mathbb Z_h^2)} = 0.
\end{equation}

We are now ready to complete the proof. We first note that
\begin{equation}
(\mathbb D_{m,h} \oplus \mathbf 0_h - z)^{-1} = (\mathbb D_{m,h} - z)^{-1} \oplus \left(-\frac 1 z\right).
\end{equation}
Now we observe that
\begin{equation} \label{eq:nonoco-4}
(\mathbb D_{m,h} - z) P_h u_h
= \frac 8{\pi^2} (m-z)\,(u_h)_h + \frac 8{\pi^2} \left(\begin{matrix} 0 \\ 
\widetilde{R}_h \end{matrix}\right) 
+ (\mathbb D_{m,h} - z) 
\left(\begin{matrix} R_h \\ 0 \end{matrix}\right),
\end{equation}
so, 
applying (\ref{eq:nonoco-3}) to the $(u_h)_h$ on the right hand side of (\ref{eq:nonoco-4}) 
and multiplying the both sides of  (\ref{eq:nonoco-4}) by
$\frac{1}{m-z}(\mathbb D_{m,h} - z)^{-1}$,
 we infer that
\begin{align*}
(\mathbb D_{m,h} - z)^{-1} P_h u_h 
&= \frac 1{m-z}\,P_h u_h - 
\frac 1{m-z}\,(\mathbb D_{m,h} - z)^{-1} 
\frac 8 {\pi^2}\left(\begin{matrix} 0 \\ \widetilde{R}_h  \end{matrix}\right) 
- \frac 1{m-z} 
\left(\begin{matrix} R_h \\ 0 \end{matrix}\right)
\\
&\qquad + (\mathbb D_{m,h} - z)^{-1} \left(\begin{matrix} R_h \\ 0 \end{matrix}\right).
\end{align*}
As $\|(\mathbb D_{m,h} - z)^{-1}\|_{\mathbf B(L^2(\mathbb Z_h^2)^2)} \le 
1/ |\mathfrak{Im}\,z |$, 
it follows that
\begin{align*}
&\|(\mathbb D_{m,h} - z)^{-1} P_h u_h - 
\frac 1{m-z}\,P_h u_h\|_{L^2(\mathbb R^2)^2}
\\
&\;\, \le \frac{8}{|m-z|\,  |\mathfrak{Im}\,z | \,\pi^2}\,
\| \widetilde{R}_h \|_{L^2(\mathbb Z_h^2)}
 + \left(\frac 1{|m-z|} + \frac{1}{|\mathfrak{Im}\,z | } \right) 
 \| R_h \|_{L^2(\mathbb Z_h^2)}
\rightarrow 0
\;\,  (h\rightarrow 0).
\end{align*}
Consequently,
\begin{align*}
\|u_h&\|_{L^2(\mathbb R^2)^2}\,\|(\mathbb D_{m,h} \oplus \mathbf 0_h - z)^{-1}
 - (\mathbb D_m - z)^{-1}\|_{\mathbf B(L^2(\mathbb R^2)^2)}
\\
&\ge \|(\mathbb D_{m,h} \oplus \mathbf 0_h - z)^{-1} u_h 
- (\mathbb D_m - z)^{-1} u_h\|_{L^2(\mathbb R^2)^2}
\\
&\ge \|(\mathbb D_{m,h} - z)^{-1} P_h u_h  
-z^{-1}(1 - P_h) u_h\|_{L^2(\mathbb R^2)^2} - 
\|(\mathbb D_m - z)^{-1} u_h\|_{L^2(\mathbb R^2)^2}
\\
&= \sqrt{\|(\mathbb D_{m_h} - z)^{-1} P_h u_h\|_{L^2(\mathbb R^2)^2}^2 + 
|z^{-1}|  \, \| (1 - P_h) u_h\|_{L^2(\mathbb R^2)^2}^2} + o(1)
\\
&\ge \frac 1{|m-z|}\,\|u_h\|_{L^2(\mathbb R^2)^2} + o(1) \qquad (h\rightarrow 0).
\end{align*}
This inequality implies that
\begin{equation}
\liminf_{h\to 0}
\Vert 
({\mathbb D}_{m,h} \oplus \mathbf 0_h - z )^{-1}  
-  ({\mathbb D}_{m} - z)^{-1}
 \, \Vert_{\mathbf B (L^2(\Rtw)^2)} \ge \frac 1 {|m - z|}.
\end{equation}

If we replace $u_h = \left(\begin{matrix}y_h \\ 0 \end{matrix}\right)$ with $v_h = \left(\begin{matrix}0 \\y_h  \end{matrix}\right)$
and make the similar arguments as above, we get
\begin{equation}
\liminf_{h\to 0}
\Vert 
({\mathbb D}_{m,h} \oplus \mathbf 0_h - z )^{-1}  
-  ({\mathbb D}_{m} - z)^{-1}
 \, \Vert_{\mathbf B (L^2(\Rtw)^2)} \ge \frac 1 {|m + z|}.
\end{equation}
We now arrived at the inequality (\ref{eq:nonoco-0}).
$\blacksquare$

\begin{rem}\label{rem:monodiff}
The lack of norm resolvent convergence shown in Theorem \ref{thm:nonoco} is closely related to the fact that, unlike
the continuous Dirac operator, the discrete Dirac operator does not control the gradient: while a simple calculation
shows that
$\|\nabla u\|_{L^2(\mathbb R^2)^2} \le \|\mathbb D_{m} u\|_{L^2(\mathbb R^2)^2}$ for all $u \in H^1(\mathbb R^2)^2$,
there is no constant $C > 0$ such that $\sqrt{\|\partial_{1,h} u\|_{\ell^2(\mathbb Z_h^2)^2}^2 + \|\partial_{2,h} u\|_{\ell^2(\mathbb Z_h^2)^2}^2} \le C \|\mathbb D_{m,h} u\|_{\ell^2(\mathbb Z_h^2)^2}$.
This, in turn, is connected to the fact that the Liouville theorem does not hold in discrete complex analysis. Indeed,
the function
$y:\mathbb Z + i \mathbb Z \rightarrow \mathbb Z + i \mathbb Z$,
$y(n_1 + i n_2) = i^{n_1 - n_2}$ satisfies the discrete Cauchy-Riemann equation
$(\partial_{1,1} + i \partial_{2,1}) y = 0$
in the whole lattice of Gaussian integers (and thus is `monodiffric', see \cite{Isaacs}, \cite{Kurowski}) and is
bounded, but not constant.
\par
The functions $y_h$ in the proof of Theorem \ref{thm:nonoco} arise from this function $y$ by a natural extension to
all of $\mathbb R^2 \cong \mathbb C$, scaling to the lattice with spacing $h$ and multiplication with a suitable
Gaussian to place the functions into Schwartz space.
\end{rem}


\vspace{10pt}

\section{Strong resolvent convergence  of ${\mathbb D}_{m,h}+ V_h$ }\label{sec:main1}

In this section, we shall discuss the continuum  Dirac operators 
\begin{equation} \label{eq:diracV-0}
{\mathbb D}_{m}+ V =  -i \sigma_1 \frac{\partial}{\partial x_1} 
-i\sigma_2  \frac{\partial}{\partial x_2} 
+ m \sigma_3 + V(x)
\;\text{  in }
L^2(\mathbb R^2)^2,
\end{equation}
where  $V$ is a complex $2\times 2$ matrix-valued potential.  More precisely, we make the following   

\vspace{10pt}
\noindent
{\bf Assumption (V).}
$V : \mathbb R^2 \rightarrow \mathbb C^{2\times 2}$ is a matrix-valued function each element of which  is a bounded and
uniformly continuous function. 

\vspace{10pt}

\medskip
\begin{rem}
It is apparent that electro-magnetic Dirac operators 
$$\sigma \cdot (-i \nabla - a(x)) + m \beta + q(x)$$ 
can
be written in the form (\ref{eq:diracV-0}). Indeed, one can take $V$ to be 
$-\sigma \cdot a(x)  + q(x)$.
\end{rem}

\medskip
We note that  $V$ can be decomposed into its Hermitian and skew-Hermitian parts,
$V = V_{\mathfrak R} + i V_{\mathfrak I}$, where
\begin{equation}\label{eq:Vsplit}
V_{\mathfrak R}(x) = \dfrac{V(x) + V(x)^*}{2}, \qquad i V_{\mathfrak I}(x) = \dfrac{V(x) -V(x)^*}{2}
 \qquad (x \in \mathbb R^2).
\end{equation}
It is evident that under the assumption (V), the operator $V$ of multiplication with the matrix-valued function $V$
is a bounded operator in  $L^2(\Rtw)^2$  and that 
the operator ${\mathbb D}_{m}+ V$ is well-defined. 
In particular, if $V_{\mathfrak I}=0$,
then ${\mathbb D}_{m}+ V$ is a self-adjoint operator in  $L^2(\Rtw)^2$
with  domain $H^1(\Rtw)^2$. 

In analogy to  (\ref{eq:diracV-0}), we consider the discrete Dirac operator $\Dh + V_h$ in 
 $L^2(\Zh^2)^2$,  where $\Dh$ is the operator introduced
in (\ref{eq:d-dirac2})  and $V_h$ is the operator of multiplication by 
\begin{equation}\label{eq:V0}
V_h(x) =  \sumZtw V(hn) \charnh  \!(x)  \qquad (x\in\Rtw).  
\end{equation}
It is clear that $\Dh + V_h$  is a bounded operator in 
$L^2(\Zh^2)^2$. 
In the same manner as in (\ref{eq:Vsplit}), we split $V_h = V_{\mathfrak{R}, h} + iV_{\mathfrak{I}, h}$ .

\vspace{8pt}
\begin{thm}\label{thm:V} 
 Under the assumption (V),
  
 \vspace{4pt}
\noindent
{\rm (i)}
 both the spectra $\sigma({\mathbb D}_{m,h} + V_h)$ and 
 $\sigma({\mathbb D}_{m} + V)$  are subsets of the strip
 $$
\big\{ \, z \in \mathbb C \, \big| \,  |{\mathfrak{Im }}\, z | 
 \le 
 \sup_{x\in \Rtw} \Vert V_{\mathfrak I}(x) \Vert_{\mathbf B (\Ctw)} \, \big\} ; 
 $$
 
 \vspace{4pt}
\noindent
{\rm (ii)}
for  $z$ with $ |{\mathfrak{Im }}\, z |  > \displaystyle{\sup_{x\in \Rtw}}\Vert V_{\mathfrak I} (x)\Vert_{\mathbf B (\Ctw)}$
\begin{equation}\label{eq:resl-eq3}
 \mathop{\hbox{\rm s-lim}}_{h\to 0} 
\big\{
 ({\mathbb D}_{m,h}+ V_h - z )^{-1}\!\oplus  \mathbf 0_h
 \big\} 
 =  
({\mathbb D}_{m} +V- z )^{-1}
\; \mbox{ in  } \,  L^2(\mathbb R^2)^2. 
\end{equation}
\end{thm}

\vspace{8pt}
\noindent
 Note that in the self-adjoint case $V_{\mathfrak I} = 0$,
 (\ref{eq:resl-eq3})  holds for
 $z \in \mathbb C \setminus \mathbb R$ .

We prepare the proof of Theorem \ref{thm:V} by providing the following auxiliary statements. 

\begin{lem} \label{lem:V0}
Suppose that (V)  holds. Then
$$
\sigma({\mathbb D}_{m} +V)
\subset
\big\{ \, z \in \mathbb C \, \big| \,  |{\mathfrak{Im }}\, z | 
 \le 
 \sup_{x\in \Rtw}\Vert V_{\mathfrak I}(x) \Vert_{\mathbf B (\Ctw)} \, \big\}.
 $$
 Moreover,
 \begin{equation}\label{eq:resl-eq2}
( {\mathbb D}_{m} +V -z )^{-1} = ({\mathbb D}_{m} +V_{\mathfrak{R}} -z )^{-1}
\big \{
 I_{L^2(\Rtw)^2}+ iV_{\mathfrak{I}} \, ({\mathbb D}_{m} +V_{\mathfrak{R}} -z )^{-1}
 \big\}^{-1}
 \end{equation}
for all $z$ with $ |{\mathfrak{Im }}\, z |  > \displaystyle{\sup_{x\in \Rtw}}
  \Vert V_{\mathfrak I}(x) \Vert_{\mathbf B (\Ctw)}$. 
  \end{lem}

 \noindent
 {\bf Proof.}
 Let 
 $ |{\mathfrak{Im }}\, z |  > \displaystyle{\sup_{x\in \Rtw}}
  \Vert V_{\mathfrak I}(x) \Vert_{\mathbf B (\Ctw)}$.
 Since ${\mathbb D}_{m} +V_{\mathfrak{R}}$ is self-adjoint, 
 it  follows that $z \in \rho({\mathbb D}_{m} +V_{\mathfrak{R}})$,
 the resolvent set of ${\mathbb D}_{m} +V_{\mathfrak{R}}$. This enables us to write
 \begin{equation}\label{eq:resl-eq1}
 {\mathbb D}_{m} +V -z  =\big \{
 I_{L^2(\Rtw)}+ iV_{\mathfrak{I}}\, ({\mathbb D}_{m} +V_{\mathfrak{R}} -z )^{-1}
 \big\}
 ({\mathbb D}_{m} + V_{\mathfrak{R}} -z ).
 \end{equation}
 Since 
$\Vert V_{\mathfrak{I}} \Vert_{\mathbf B (L^2(\Rtw)^2)}
\le
{\sup_{x\in \Rtw}}\Vert V_{\mathfrak I}(x) \Vert_{\mathbf B (\Ctw)} $,
 we see that 
$$\Vert V_{\mathfrak I} ({\mathbb D}_{m} +V_{\mathfrak{R}} -z )^{-1}\Vert_{\mathbf B (L^2(\Rtw)^2)} <1,$$ so the operator on  the right hand side of
 (\ref{eq:resl-eq1}) is invertible in  $ L^2(\mathbb R)^2$,
 and therefore $z \in \rho({\mathbb D}_{m} +V)$.
$\blacksquare$

 \vspace{8pt}
 In the same manner as in the proof of Lemma \ref{lem:V0}, one can prove a similar  statement  for
 ${\mathbb D}_{m, h} +V_h$. Recall that ${\mathbb D}_{m, h} +V_h$ is
 a bounded operator acting in  $L^2(\Zhtw)^2$.

  \begin{lem} \label{lem:V1}
Under assumption (V), 
$$
\sigma({\mathbb D}_{m, h} +V_h)
\subset
\big\{ \, z \in \mathbb C \, \big| \,  |{\mathfrak{Im }}\, z | 
 \le 
 \sup_{x\in \Rtw}\Vert V_{\mathfrak I}(x) \Vert_{\mathbf B (\Ctw)} \, \big\}.
 $$
 Moreover,
 \begin{equation}\label{eq:resl-eq4}
\begin{aligned}
( {\mathbb D}_{m,h} &+V_h -z )^{-1} \\ &= ({\mathbb D}_{m,h} +V_{\mathfrak{R},h} -z )^{-1}
\big \{
 I_{L^2(\Zhtw)^2}+ i V_{\mathfrak{I},h} \, ({\mathbb D}_{m,h} +V_{\mathfrak{R},h} -z )^{-1}
 \big\}^{-1}
\end{aligned}
 \end{equation}
for all  $z$ with $|{\mathfrak{Im }}\, z |  > \displaystyle{\sup_{x\in \Rtw}}
  \Vert V_{\mathfrak I}(x) \Vert_{\mathbf B (\Ctw)}$.
\end{lem}

\vspace{10pt}
\begin{lem} \label{lem:V2}
 Under the assumption {\rm ($\!$}V\,{\rm)},
$V_{h}\oplus  \mathbf 0_h \,(=V_h P_h)  \to  V $ strongly in  $L^2(\Rtw)^2$. 
\end{lem}

\noindent
{\bf Proof.}  
In this proof, we distinguish the multiplication operator
 $V_{h}$ in  $L^2(\Rtw)^2$
from the embedded version of the multiplication operator
$V_{h}$ in $L^2(\Zhtw)^2$, denoted by
$V_{h}\oplus  \mathbf 0_h$.

Let $\varphi\in L^2(\Rtw)^2$. 
 In view of the fact that
$\{V_{h}\oplus  \mathbf 0_h \} \varphi = V_{h} P_h \varphi$,
 we infer that
\begin{equation*}
\begin{aligned}
\big\Vert \big[\{V_{h}\oplus  \mathbf 0_h \} - V\big] \varphi  \Vert_{L^2(\Rtw)^2}
&\le
\left\Vert V_h\,(P_h \varphi - \varphi) \right\Vert_{L^2(\Rtw)^2}
 + \left\Vert (V_{h} - V)\varphi  \right\Vert_{L^2(\Rtw)^2}   \\
& \le
 \left\{ \sup_{x\in \Rtw}\Vert V(x) \Vert_{\mathbf B (\Ctw)} \right\} 
 \left\Vert P_h \varphi - \varphi \right\Vert_{L^2(\Rtw)^2} \\
& \qquad +
 \sup_{n\in \Ztw}\big\{ \sup_{x\in I_{\!n\!,h}}\Vert V(x) -  V(hn)\Vert_{\mathbf B (\Ctw)} \big\} 
 \big\Vert \varphi \big\Vert_{L^2(\Rtw)^2} .
\end{aligned}                                                                                                                                                                                                                                                                                                                                                      
\end{equation*}
By virtue of  Lemma \ref{lem:Ph}  and the assumption that the function $V$ is bounded and uniformly continuous, it follows that
$V_{h}\oplus \mathbf 0_h \to  V $
strongly in $L^2(\Rtw)^2 $ as $h\to 0$. 
$\blacksquare$

\medskip
\begin{rem}
As shown in Lemma \ref{lem:V2}, $V_{h}\oplus  \mathbf 0_h$ converges to $V$ strongly, but not in the 
operator norm unless $V\equiv 0$.
Indeed, if $V \not\equiv 0$, then there is some open subset $\Omega\subset\mathbb{R}^2$, some $v\in\mathbb{C}^2$ with
$\|v\|_{\mathbb{C}^2} = 1$ and a constant $C > 0$ such that $\|V(x) v\|_{\mathbb{C}^2} \ge $ for all $x\in\Omega$.
Let $h > 0$ be so small that for some $n\in\mathbb{Z}^2$, $I_{n,h} \subset\Omega$, and set
\begin{equation*}
\varphi(x) = \chi_{I_{n,h}}(x) \prod_{j=1}^2 (x_j - h(n+\frac 1 2))\,v \qquad (x\in\mathbb{R}^2).
\end{equation*}
Then
\begin{align*}
\|(V_h \oplus 0 &- V)\varphi\|_{L^2(\mathbb{R}^2)} = \|V_h P_h \varphi - V \varphi\|_{L^2(\mathbb{R}^2)}
= \|V \varphi\|_{L^2(\mathbb{R}^2)}
\\
&= 
\Big\{\int_{I_{n,h}} \|V(x) v\|_{\mathbb{C}^2}^2 \prod_{j=1}^2 |x_j - h(n_j+\frac 1 2)|^2\,d x\Big\}^{1/2}
\ge C \|\varphi\|_{L^2(\mathbb{R}^2)}.
\end{align*}
Therefore $\|V_h\oplus 0 -V\|_{\mathbf{B}(L^2(\mathbb{R}^2))} \ge C > 0$ for all sufficiently small $h > 0$.

This remark shows that even at the level of the potential operator $V$, we cannot expect norm convergence, which is
slightly counterintuitive, as $V_h$,  as a function, does converge in $\|\cdot\|_\infty$ norm to $V$. 
\end{rem}

\medskip
Applying the above lemma to $V$ and to $V^*$ and using (\ref{eq:Vsplit}), 
we obtain the following convergence
results for the Hermitian and skew-Hermitian parts separately.

\begin{cor} \label{cor:V2}
Suppose that (V) is verified. Then 
$V_{\mathfrak{R},h}\oplus  \mathbf 0_h \to  V_{\mathfrak{R}}$ strongly in  $L^2(\Rtw)^2$
and 
 $V_{\mathfrak{I},h}\oplus  \mathbf 0_h \to  V_{\mathfrak{I}}$
  strongly in  $L^2(\Rtw)^2$. 
\end{cor}

 \medskip
Furthermore, we shall use the following two abstract lemmas. We omit the quite straightforward proof of the first
of them.

\begin{lem}\label{lem:abs1}
Let ${\mathcal H}$ be a Hilbert space.  Let $S_h$ and $T_h$
belong to  ${\mathbf B}(\mathcal H)$ for each $h>0$, 
and  suppose that
$S_h$ and $T_h$ strongly converge to $S$ and $T\in {\mathbf B}(\mathcal H)$ respectively as $h\to 0$.
If  \,$\sup_{h > 0}  \Vert S_h \Vert_{{\mathbf B}(\mathcal H)} < \infty$, 
then 
$S_h T_h$ strongly converges to $ST$ as $h\to 0$.
\end{lem}

\begin{lem}\label{lem:abs0}
Let ${\mathcal H}$ be a Hilbert space.  Suppose that ${\mathcal H}$ has an orthogonal decomposition
${\mathcal H}={\mathcal X}_h \oplus {\mathcal X}_h^{\perp}$
for each $h>0$  and that
the orthogonal projection $P_h$ onto ${\mathcal X}_h$
strongly converges to $I_{\mathcal H}$
 as $h\to 0$.
 Let  $A_h$, for each $h > 0$,  and $A$ be  invertible operators in $\mathcal X_h$ and  in $\mathcal H$ respectively such that
 $A_h\oplus 0$ strongly converges to $A$ as $h\to 0$.  
 If $\sup_{h>0}\Vert A_h^{-1} \Vert_{\mathbf B({\mathcal X}_h)} < \infty$,
 then 
 $A_h^{-1}\oplus 0$ strongly converges to $A^{-1}$ as $h\to 0$.
\end{lem}

 \noindent
 {\bf Proof.}  
 Let $\varphi \in \mathcal H$. By hypothesis, we see that
 \begin{equation*}
 (A_h\oplus \mathbf 0_h) A^{-1} \varphi  \to A A^{-1}\varphi  = \varphi \; \mbox{ in } \mathcal H
 \end{equation*}
 as $h \to 0$.
 Hence we obtain
\begin{align*}
\|A^{-1}\varphi &- (A_h^{-1}\oplus \mathbf 0_h)\varphi\|_{\mathcal H} \\
&\le \|(I_{\mathcal H} - P_h) A^{-1}\varphi\|_{\mathcal H} 
 + \|(A_h^{-1}\oplus \mathbf 0_h)\,((A_h\oplus \mathbf 0_h) A^{-1}\varphi - \varphi)\|_{\mathcal H}
  \rightarrow 0 \quad (h \rightarrow 0)
\end{align*}
by the uniform boundedness of $A_h^{-1}$. 
 $\blacksquare$

\medskip
\noindent
{\bf Proof of Theorem \ref{thm:V}.}
Statement (i) was shown in Lemma \ref{lem:V0} and Lemma \ref{lem:V1}. For (ii),
let $z \in \mathbb C$, $|\mathop{\mathfrak{Im}} z| > \sup_{x\in\Rtw} \|V_{\mathfrak{I}}\|_{\mathbf B(\Ctw)}$. Then
Lemmas \ref{lem:V2}, \ref{lem:abs1}, and Theorem \ref{thm:x}, together with the fact that
\begin{equation}
\Vert V_{h}\oplus \mathbf 0_h \Vert_{\mathbf B(L^2(\Rtw)^2)}
\le 
\sup_{x\in\Rtw}  \Vert V(x) \Vert_{\mathbf B(\Ctw)}  < \infty
\end{equation}
imply that
\begin{equation}
\big\{V_{h} \,({\mathbb D}_{m,h}  -z )^{-1} \big\} \oplus \mathbf 0_h 
=
\big\{V_{h}  \oplus \mathbf 0_h \big\}
\big\{({\mathbb D}_{m,h}  -z )^{-1}  \oplus \mathbf 0_h \big\}  
\to  
\, V ({\mathbb D}_{m}  -z )^{-1}
\end{equation}
strongly in $L^2(\Rtw)^2$ as $h\to 0$,
so also
\begin{equation}\label{eq:V2}
\big\{I_{L^2(\Zhtw)}+ V_{h} \,({\mathbb D}_{m,h}  -z )^{-1} \big\} \oplus \mathbf 0_h 
\to
I_{L^2(\Rtw)}+ V ({\mathbb D}_{m}  -z )^{-1}  
\end{equation}
strongly in $L^2(\Rtw)^2$ as $h\to 0$.

Since $z$ lies in the resolvent set of both ${\mathbb D}_{m,h}$ and ${\mathbb D}_{m,h} + V_{h}$
(see Lemma \ref{lem:V1}), we 
see that the right hand (and therefore the left hand) side of 
\begin{equation}
I_{L^2(\Zhtw)^2}+ V_{h} ({\mathbb D}_{m,h}  -z )^{-1} =
\,({\mathbb D}_{m,h} + V_{h}  -z)( {\mathbb D}_{m,h}   -z)^{-1} 
\end{equation}
is invertible in $L^2(\Zhtw)^2$, so
\begin{equation*}
({\mathbb D}_{m,h} + V_{h}  -z)^{-1}=
( {\mathbb D}_{m,h}   -z)^{-1} 
\big\{ I_{L^2(\Zhtw)^2}+ V_{h} ({\mathbb D}_{m,h}  -z )^{-1} \big\}^{-1}.
\end{equation*}
Now, in order to apply Lemma \ref{lem:abs0} with 
$$
A_h=
 I_{L^2(\Zhtw)^2}+ V_{h} ({\mathbb D}_{m,h}  -z )^{-1}  \mbox{ in }L^2(\Zhtw)^2
$$
and
$$
A=
I_{L^2(\Rtw)^2}+ V ({\mathbb D}_{m}  -z )^{-1}   \mbox{ in }L^2(\Rtw)^2,
$$
we require uniform boundedness of $A_h^{-1}$. 
To this end, we note that
one can write 
\begin{equation}\label{eq:Ainv}
A_h^{-1}=  I_{L^2(\Zhtw)^2} - V_{h} ({\mathbb D}_{m,h} + V_{h}  -z )^{-1},
\end{equation}
and analogously
\begin{equation*}
 \left(I_{L^2(\Zhtw)^2} + i V_{\mathfrak{I},h} (\mathbb D_{m,h} + V_{\mathfrak{R},h} - z)^{-1} \right)^{-1}
= I_{L^2(\Zhtw)^2} - i V_{\mathfrak{I},h} (\mathbb D_{m,h} + V_h - z)^{-1};
\end{equation*}
so we find, using (\ref{eq:resl-eq4}),
\begin{align*}
&\|(\mathbb D_{m,h} + V_h - z)^{-1}\|_{\mathbf B(L^2(\Zhtw)^2)} \\
&= \left\|(\mathbb D_{m,h} + V_{\mathfrak{R},h} - z)^{-1}\,\left(I_{L^2(\Zhtw)^2} + i V_{\mathfrak{I},h} (\mathbb D_{m,h} + V_{\mathfrak{R},h} - z)^{-1} \right)^{-1}\right\|_{\mathbf B(L^2(\Zhtw)^2)} \\
&\le \|(\mathbb D_{m,h} + V_{\mathfrak{R},h} - z)^{-1}\|_{\mathbf B(L^2(\Zhtw)^2)}
  \|I_{L^2(\Zhtw)^2} - i V_{\mathfrak{I},h} (\mathbb D_{m,h} + V_h - z)^{-1}\|_{\mathbf B(L^2(\Zhtw)^2)} \\
&\le \frac 1 {|\mathop{\mathfrak{Im}} z|} \left(1 + \sup_{x\in\Rtw} \|V_{\mathfrak{I}}(x)\|_{\mathbf B(\Ctw)}
  \|(\mathbb D_{m,h} + V_h - z)^{-1}\|_{\mathbf B(L^2(\Zhtw)^2)} \right),
\end{align*}
giving
\begin{equation*}
 \|(\mathbb D_{m,h} + V_h - z)^{-1}\|_{\mathbf B(L^2(\Zhtw)^2)} \le \frac 1 {|\mathop{\mathfrak{Im}} z| - \sup_{x\in\Rtw} \|V_{\mathfrak{I}}(x)\|_{\mathbf B(\Ctw)}}
\end{equation*}
and further by (\ref{eq:Ainv})
\begin{align*}
 \|A_h^{-1}\|_{\mathbf B(L^2(\Zhtw)^2)} &\le 1 + \sup_{x\in\Rtw} \|V_{\mathfrak{I}}(x)\|_{\mathbf B(\Ctw)} \|(\mathbb D_{m,h} + V_h - z)^{-1}\|_{\mathbf B(L^2(\Zhtw)^2)} \\
 &\le \frac {|\mathop{\mathfrak{Im}} z|}{|\mathop{\mathfrak{Im}} z| - \sup_{x\in\Rtw} \|V_{\mathfrak{I}}(x)\|_{\mathbf B(\Ctw)}}
\end{align*}
for all $h > 0$.
Thus we can conclude, with the help of (\ref{eq:V2}) and Lemma \ref{lem:abs0}, that
 \begin{equation}\label{eq:V3}
 \big\{ I_{L^2(\Zhtw)^2}+ V_{h} ({\mathbb D}_{m,h}  -z )^{-1} \big\}^{-1}\oplus  \mathbf 0_h
\to
\big\{ I_{L^2(\Rtw)^2}+ V ({\mathbb D}_{m}  -z )^{-1} \big\}^{-1}
 \end{equation}
strongly in $L^2(\Rtw)^2$.
 Noting that 
\begin{equation*}
\Vert  ({\mathbb D}_{m,h}   -z)^{-1} \oplus \mathbf 0_h \Vert_{L^2(\Rtw)^2 } 
\le 
\frac{1}{ \, |\mathfrak{Im} z| \,}
\end{equation*}
for all $h > 0$,
we see from Lemma \ref{lem:abs1}, Theorem \ref{thm:x} and (\ref{eq:V3}) that
\begin{equation*}
({\mathbb D}_{m,h} + V_{h}  -z)^{-1}\oplus  \mathbf 0_h  \to
({\mathbb D}_{m}   -z)^{-1}
\big\{ I_{L^2(\Rtw)}+ V ({\mathbb D}_{m}  -z )^{-1} \big\}^{-1} 
\end{equation*}
strongly in $L^2(\Rtw)^2$.
This completes the proof. $\blacksquare$


  \end{document}